\documentclass[11pt,a4paper]{article}
\usepackage{subfigure}
\usepackage{jheppub}
\usepackage{epsfig}
\usepackage{amssymb,amsmath}
\usepackage{graphicx}
\usepackage{color}

\title{An analysis of the impact of LHC Run I proton-lead data on nuclear parton densities}

\author[a]{N\'estor Armesto, }
\author[a,b,c]{Hannu Paukkunen, }
\author[a]{Jos\'e Manuel Pen\'in, } 
\author[a]{Carlos A. Salgado } 
\author[a]{and P\'ia Zurita}

\affiliation[a]{Departamento de F\'\i sica de Part\'\i culas and IGFAE, Universidade de Santiago de
Compostela, E-15782 Galicia, Spain}
\affiliation[b]{Department of Physics, University of Jyv\"askyl\"a, P.O. Box 35, FI-40014 University of Jyv\"askyl\"a, Finland}
\affiliation[c]{Helsinki Institute of Physics, P.O. Box 64, FI-00014 University of Helsinki, Finland}

\emailAdd{nestor.armesto@usc.es}
\emailAdd{hannu.paukkunen@jyu.fi}
\emailAdd{jmanpen@gmail.com}
\emailAdd{carlos.salgado@usc.es}
\emailAdd{pia.zurita@usc.es} 	

\abstract{
We report on an analysis of the impact of available experimental data on hard processes in proton-lead collisions during Run I at the Large Hadron Collider on nuclear modifications of parton distribution functions. Our analysis is restricted to the EPS09 and DSSZ global fits. The measurements that we consider comprise production of massive gauge bosons, jets, charged hadrons and pions. This is the first time a study of nuclear PDFs includes this number of different observables. The goal of the paper is twofold: i) checking the description of the data by nPDFs, as well as the relevance of these nuclear effects, in a quantitative manner; ii) testing the constraining power of these data in eventual global fits, for which we use the Bayesian reweighting technique. We find an overall good, even too good, description of the data, indicating that more constraining power would require a better control over the systematic uncertainties and/or the proper proton-proton reference from LHC Run II. Some of the observables, however, show sizable tension with specific choices of proton and nuclear PDFs. We also comment on the corresponding improvements on the theoretical treatment. 
}

\begin{document}

\maketitle

\section{Introduction}\label{sec:intro}

The main physics motivations \cite{Salgado:2011wc} for the proton-lead (p-Pb) collisions at the Large Hadron Collider (LHC) were to obtain a reliable baseline for the heavy-ion measurements and to shed light on the partonic behaviour of the nucleus, particularly at small values of momentum fraction $x$. As such, this program constitutes a logical continuation of the deuteron-gold (d-Au) experiments at the Relativistic Heavy-Ion Collider (RHIC) but at significantly higher energy. The p-Pb data has, however, proved richer that initially pictured and also entailed  genuine surprises (see the review \cite{Armesto:2015ioy}). 

One of the key factors in interpreting the p-Pb data are the nuclear parton distribution functions (nPDFs) \cite{Eskola:2012rg,Paukkunen:2014nqa}. It is now more than three decades ago that, unexpectedly, large nuclear effects in deeply inelastic scattering were first found (for a review, see Ref.~\cite{Arneodo:1992wf}), which were later on shown to be factorisable into the PDFs \cite{Eskola:1998iy}. However, the amount and variety of experimental data that go into the global determinations of nPDFs has been very limited and the universality of the nPDFs has still remained largely as a conjecture --- with no clear violation found to date, however. The new experimental data from the LHC p-Pb run give a novel opportunity to further check these ideas and also provide new constraints. The aim of this paper is, on the one hand, to chart the importance of nPDFs in describing the data (both globally and separately for individual data sets) and, on the other hand, to estimate the quantitative constraints that these data render. The latter question would have traditionally required a complete reanalysis adding the new data on top of the old ones. Luckily, faster methods, collectively known as reweighting techniques, have been developed \cite{Giele:1998gw,Ball:2010gb,Ball:2011gg,Watt:2012tq,Watt:2013oha,Sato:2013ika,Paukkunen:2014zia}.

In a preceding work \cite{Armesto:2013kqa}, a specific version \cite{Watt:2012tq} of the Bayesian reweighting technique was employed to survey the potential impact of the p-Pb program on nPDFs by using pseudodata. However, at that point the reweighting method used was not yet completely understood and certain caution regarding the results has to be practiced. Along with the developments of Ref.~\cite{Paukkunen:2014zia}, we can now more reliably apply the Bayesian reweighting. Also, instead of pseudodata we can now use available p-Pb measurements. We will perform the analysis with two different sets of nPDFs (EPS09 \cite{Eskola:2009uj} and DSSZ \cite{deFlorian:2011fp}) and, in order to control the bias coming from choosing a specific set of free proton reference, we will consider two sets of proton PDFs (MSTW2008 \cite{Martin:2009iq} and CT10 \cite{Lai:2010vv}).

The paper is organised as follows: In Section \ref{sec:details} we briefly explain the Bayesian reweighting, devoting Section \ref{sec:experiments} to the observables included in the present analysis. In Section \ref{sec:results} we show the of data impact on the nPDFs, and discuss similarities and differences between the four possible PDF-nPDF combinations. Finally, in Section \ref{sec:summary} we summarise our findings.

\section{The reweighting procedure}\label{sec:details}

\subsection{The Bayesian reweighting method}\label{sec:Bayes}

The Bayesian reweighting technique \cite{Giele:1998gw,Ball:2010gb,Ball:2011gg,Watt:2012tq,Watt:2013oha,Sato:2013ika,Paukkunen:2014zia} is a tool to quantitatively determine the implications of new data within a set of PDFs. In this approach, the probability distribution $\mathcal{P}_{\rm old}(f)$ of an existing PDF set is represented by an ensemble of PDF replicas $f_k$, $k=1, \ldots, N_{\rm rep}$, and the expectation value $\langle \mathcal{O} \rangle$ and variance $\delta \langle \mathcal{O} \rangle$ for an observable $\mathcal{O}$ can be computed as
\begin{eqnarray}
\langle \mathcal{O} \rangle  & = & \frac{1}{N_{\rm rep}} \sum_{k=1}^{N_{\rm rep}} \mathcal{O} \left[ f_k \right], \label{eq:promedio}\\
\delta \langle \mathcal{O} \rangle  & = & \sqrt{\frac{1}{N_{\rm rep}} \sum_{k=1}^{N_{\rm rep}} \left( \mathcal{O} \left[ f_k \right] - \langle \mathcal{O} \rangle\right)^2 }.
\end{eqnarray}
Additional information from a new set of data $\vec y$ can now be incorporated, by the Bayes theorem, as
\begin{equation}
\mathcal{P}_{\rm new}(f) \varpropto \mathcal{P}(\vec y \vert f) \, \mathcal{P}_{\rm old}(f)\, ,
\label{eq:proba}
\end{equation}
where $\mathcal{P}(\vec y  \vert f)$ stands for the conditional probability for the new data, for a given set of PDFs. It follows that the average value for any observable depending on the PDFs becomes a weighted average:
\begin{eqnarray}
\langle \mathcal{O} \rangle_{\rm new}  & = & \frac{1}{N_{\rm rep}} \sum_{k=1}^{N_{\rm rep}} \omega_k \, \mathcal{O} \left[ f_k \right], \label{eq:Bnew} \\
\delta \langle \mathcal{O} \rangle_{\rm new}  & = & \sqrt{\frac{1}{N_{\rm rep}} \sum_{k=1}^{N_{\rm rep}} \omega_k \, \left( \mathcal{O} \left[ f_k \right] - \langle \mathcal{O} \rangle_{\rm new} \right)^2 } \, , \label{eq:Bnew2}
\end{eqnarray}
where the weights $\omega_{k}$ are proportional to the likelihood function $\mathcal{P}(\vec y \vert f)$. For PDF sets with uncertainties based on the Hessian method (with $N_{\rm eig}$ eigenvalues resulting in $2 N_{\rm eig}+1$ members) and fixed tolerance $\Delta \chi^2$ (which is the case in the present study), the functional form of the likelihood function that corresponds to a refit \cite{Paukkunen:2014zia} is
\begin{equation}
\omega_k
= \frac{\exp\left[-\chi^2_k/2\Delta \chi^2\right]}{(1/N_{\rm rep}) \sum_{k=1}^{N_{\rm rep}} 
\exp\left[-\chi^2_k/2\Delta \chi^2\right]}, \label{eq:wGK}
\end{equation}
where 
\begin{equation}
 \chi^{2}_{k} = \sum_{i,j=1}^{N_{\rm data}} \left(y_i[f_k]-y_i\right) C_{ij}^{-1} \left(y_j[f_k]-y_j\right) \, , \label{eq:chi2onlynew}
\end{equation}
and $C$ is the covariance matrix. The ensemble of PDFs required by this approach is defined by
\begin{equation}
 f_k \equiv f_{S_0} + \sum_i^{N_{\rm eig}} \left( \frac{f_{S^+_i}-f_{S^-_i}}{2} \right) R_{ik} \label{eq:replicas},
\end{equation}
where $f_{S_0}$ is the central fit, and $f_{S^\pm_i}$ are the $i$th error sets. The coefficients $R_{ik}$ are random numbers selected from a Gaussian distribution centred at zero and with variance one. After the reweighting, the values of $\chi^2$ are evaluated as
\begin{equation}
\chi^{2}_{\rm post-rw} = \sum_{i,j=1}^{N_{\rm data}} \left(\langle y_i \rangle-y_i\right) C_{ij}^{-1} \left(\langle y_j \rangle-y_j\right),
\label{eq:chi2a}
\end{equation}
where $\langle y_i \rangle$ are computed as in Eq.~(\ref{eq:Bnew}). An additional quantity in the Bayesian method is the effective number of replicas $N_{\rm eff}$, a useful indicator defined as
\begin{equation}
N_{\rm eff} \equiv \exp \left\{ \frac{1}{N_{\rm rep}} \sum_{k=1}^{N_{\rm rep}} \omega_k \log(N_{\rm rep}/\omega_k)\right\}.
\end{equation}
Having $N_{\rm eff} \ll N_{\rm rep}$ indicates that some of the replicas are doing a significantly better job in describing the data than others, and that the method becomes inefficient. In this case a very large number of replicas may be needed to obtain a converging result. In this work we have taken $N_{\rm rep}=10^4$.


\subsection{Bayesian reweighting in the linear case}\label{sec:fastBayes}

The reweighting procedure begins by first generating the replicas $f_k$ by Eq.~(\ref{eq:replicas}), which are then used to compute the observables required to evaluate the values of $\chi^{2}_{k}$ that determine the weights. In general, this involves looping the computational codes over the $N_{\rm rep}$ replicas, which can render the work quite CPU-time consuming. There is, however, a way to reduce the required time if the PDFs that we are interested in enter the computation linearly. Let us exemplify this with the process ${\rm p} + {\rm Pb} \rightarrow \mathcal{O}$. The cross section corresponding to the $k$-th replica can be schematically written as 
\begin{equation}
d\sigma_{k}= f^{\rm p} \otimes \hat{\sigma}(\mathcal{O}) \otimes f^{\rm Pb}_{k} \, ,
\end{equation}
where $\otimes$ denotes in aggregate the kinematic integrations and summations over the partonic species. If now we replace $f^{\rm Pb}_{k}$ by Eq. (\ref{eq:replicas}), we have
\begin{equation}
d\sigma_{k}= f^{\rm p} \otimes \hat{\sigma}(\mathcal{O}) \otimes \left[f^{\rm Pb}_{S_0} + \sum_{i}^{N_{\rm eig}} \left( \frac{f^{\rm Pb}_{S^{+}_i}-f^{\rm Pb}_{S^{-}_i}}{2} \right) R_{ik}\right]  \, ,
\end{equation}
which can be written as
\begin{equation}
d\sigma_{k} = d\sigma_{S_0} + \sum_{i}^{N_{\rm eig}} \frac{R_{ik}}{2} \left[ d\sigma_{S_i^+} - d\sigma_{S_i^-} \right],
\end{equation}
where $d\sigma_{S_0}$ is the cross section obtained with the central set, and $d\sigma_{S_i^\pm}$ are the cross sections evaluated with the error sets.
In this way, only $2N_{\rm eig}+1$ (31 for EPS09, 51 for DSSZ) cross-section evaluations are required (instead of $N_{\rm rep}$).

\section{Comparison with the experimental data}\label{sec:experiments}

All the data used in this work ($165$ points in total) were obtained at the LHC during Run I, in p-Pb collisions at a centre-of-mass energy $\sqrt{s}=5.02 \, {\rm TeV}$ per nucleon: $W$ from ALICE and CMS, $Z$ from ATLAS and CMS, jets from ATLAS, dijets from CMS, charged hadrons from ALICE and CMS, and pions from ALICE. Some of them are published as absolute distributions and some as ratios. We refrain from directly using the absolute distributions as they are typically more sensitive to the free proton PDFs and not so much to the nuclear modifications. In ratios of cross sections, the dependence of the free proton PDFs usually becomes suppressed. The ideal observable would be the nuclear modification $\sigma({\rm p}$-${\rm Pb})/\sigma({\rm p}$-${\rm p})$. However, no direct p-p measurement exists yet at the same centre-of-mass energy and such a reference is sometimes constructed by the experimental collaborations from their results at $\sqrt{s}=2.76 \, {\rm TeV}$ and $\sqrt{s}=7 \, {\rm TeV}$. This brings forth a non-trivial normalisation issue and, with the intention of avoiding it, we decided to use (whenever possible) ratios between different rapidity windows instead --- this situation is expected to be largely improved in the near future thanks to the reference p-p run at $\sqrt{s}=5.02 \, {\rm TeV}$ from LHC Run II. We note that, apart from the luminosity, no information on the correlated systematic uncertainties is given by the experimental collaborations. Thus, when constructing ratios of cross sections, we had no other option than adding all the uncertainties in quadrature. In the (frequent) cases where the systematic uncertainties dominate, this amounts to overestimating the uncertainties which sometimes reflects in absurdly small logarithmic likelihood, $\chi^2/N_{\rm data} \ll 1$. The fact that the information of the correlations is not available undermines the usefulness of the data to constrain the theory calculations. This is a clear deficiency of the measurements and we call for publishing the information on the correlations as is usually done in the case of p-p and p-$\overline{\rm p}$ collisions. It is also worth noting that we (almost) only use minimum bias p-Pb data. While centrality dependent data are also available, it is known that any attempt to classify event centrality results in imposing a non-trivial bias on the hard-process observable in question, see e.g. Ref.~\cite{Armesto:2015kwa}.

Note that not all PDF+nPDF combinations will be shown in the figures to limit the number of plots. Moreover, the post-reweighting results are not shown when they become visually indistinguishable from the original ones.



\subsection{Charged electroweak bosons}

Charged electroweak bosons ($W^{+}$ and $W^{-}$) decaying into leptons have been measured by the ALICE \cite{Zhu:2015kpa} and CMS \cite{Khachatryan:2015hha} Collaborations.\footnote{Also preliminary ATLAS data have been shown \cite{ATLASWprelim} and they appear consistent with the CMS results.} The theoretical values were computed at next-to-leading order (NLO) accuracy using the Monte Carlo generator MCFM \cite{Campbell:2011bn} fixing all the QCD scales to the mass of the boson.

\begin{figure}[h!]
\centering
\includegraphics[width=0.5\textwidth]{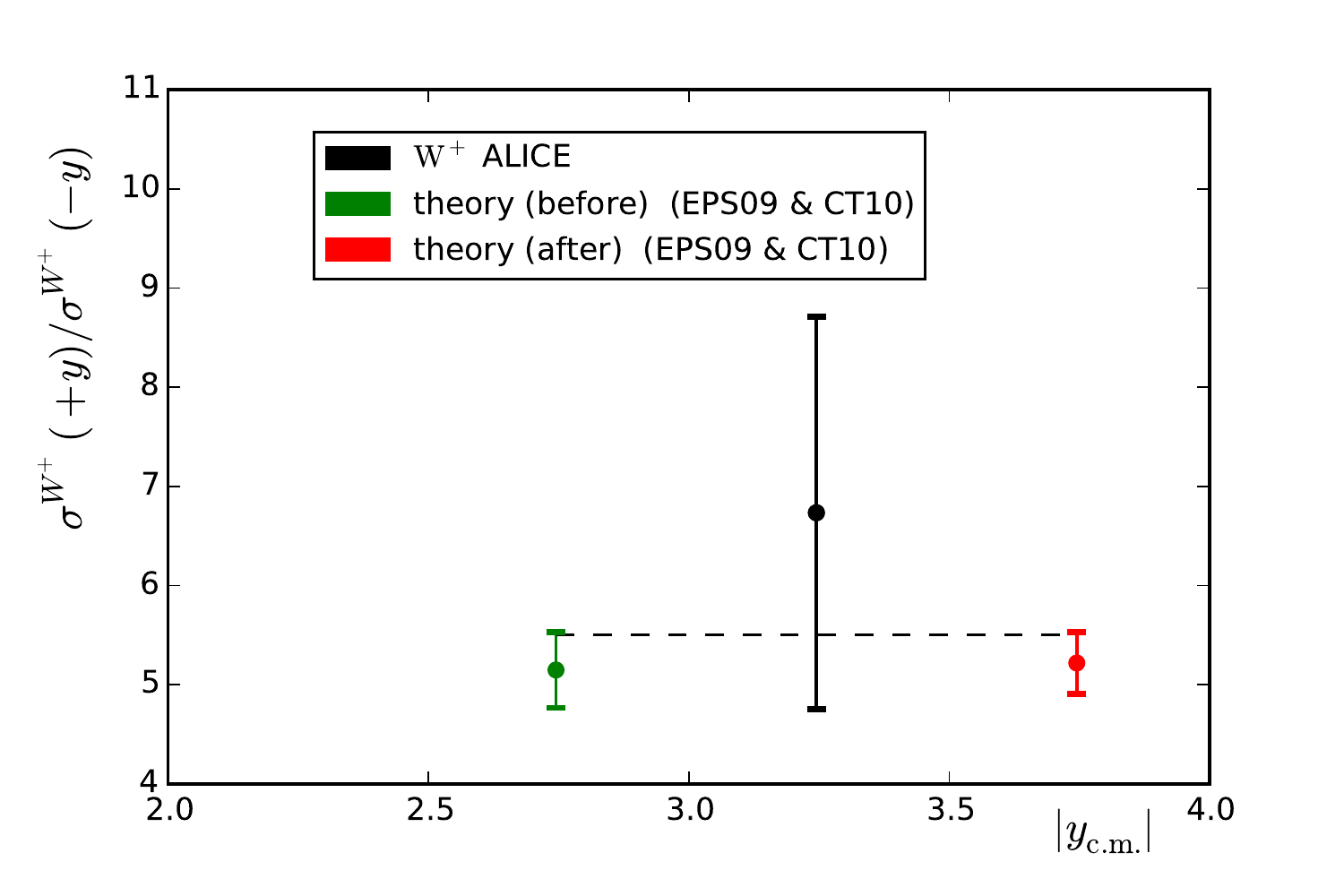} \hspace{-0.5cm}
\includegraphics[width=0.5\textwidth]{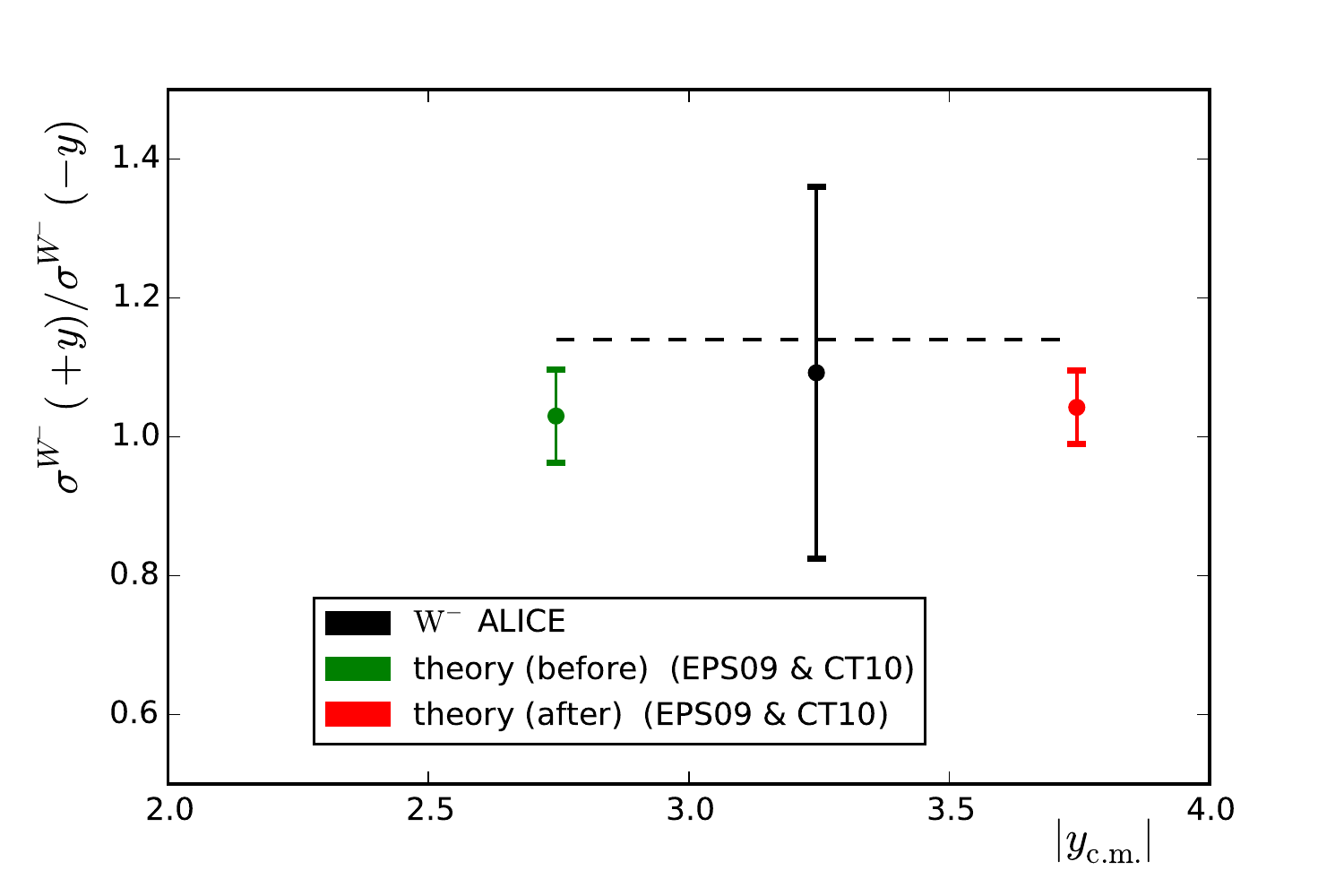}
\includegraphics[width=0.5\textwidth]{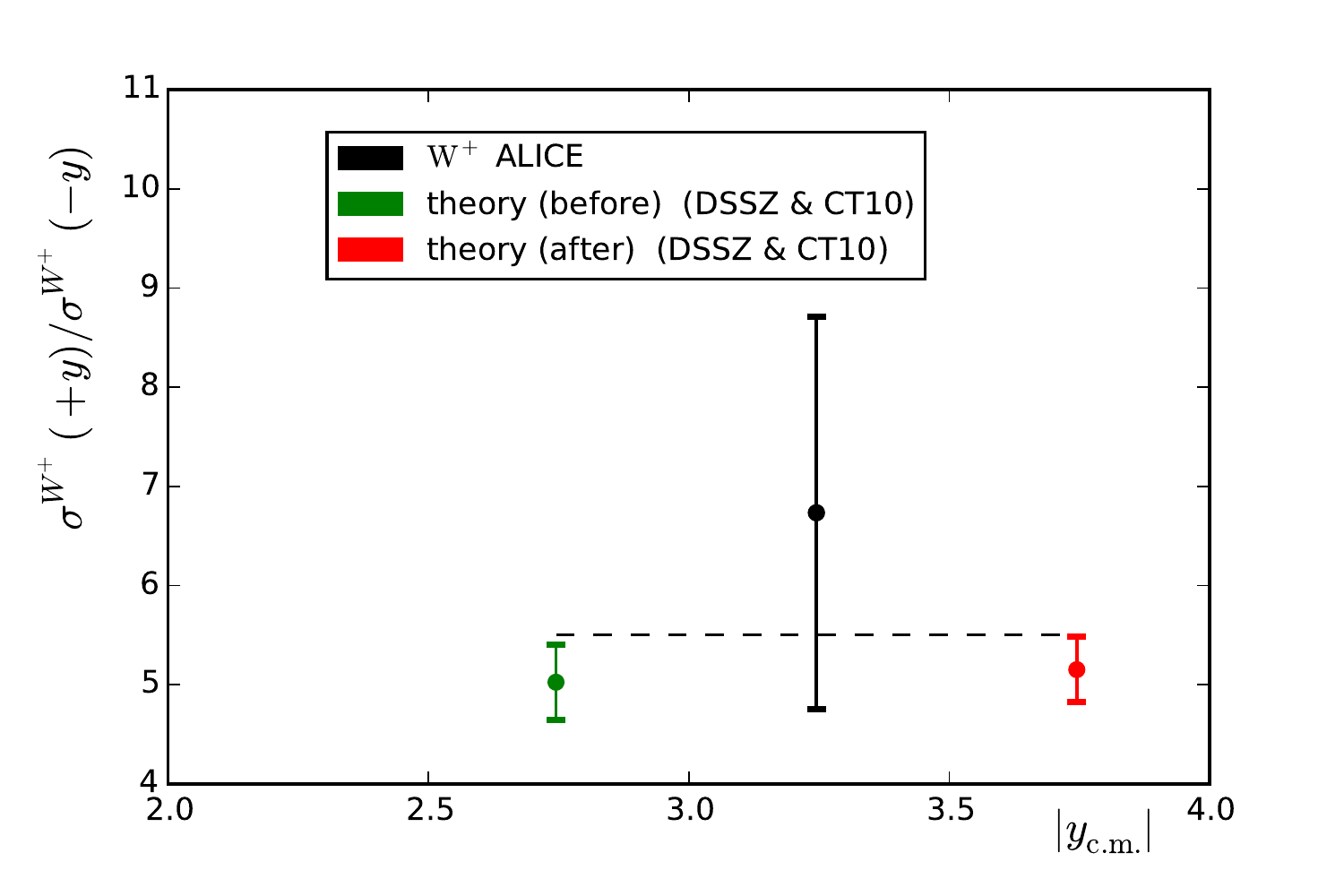} \hspace{-0.5cm}
\includegraphics[width=0.5\textwidth]{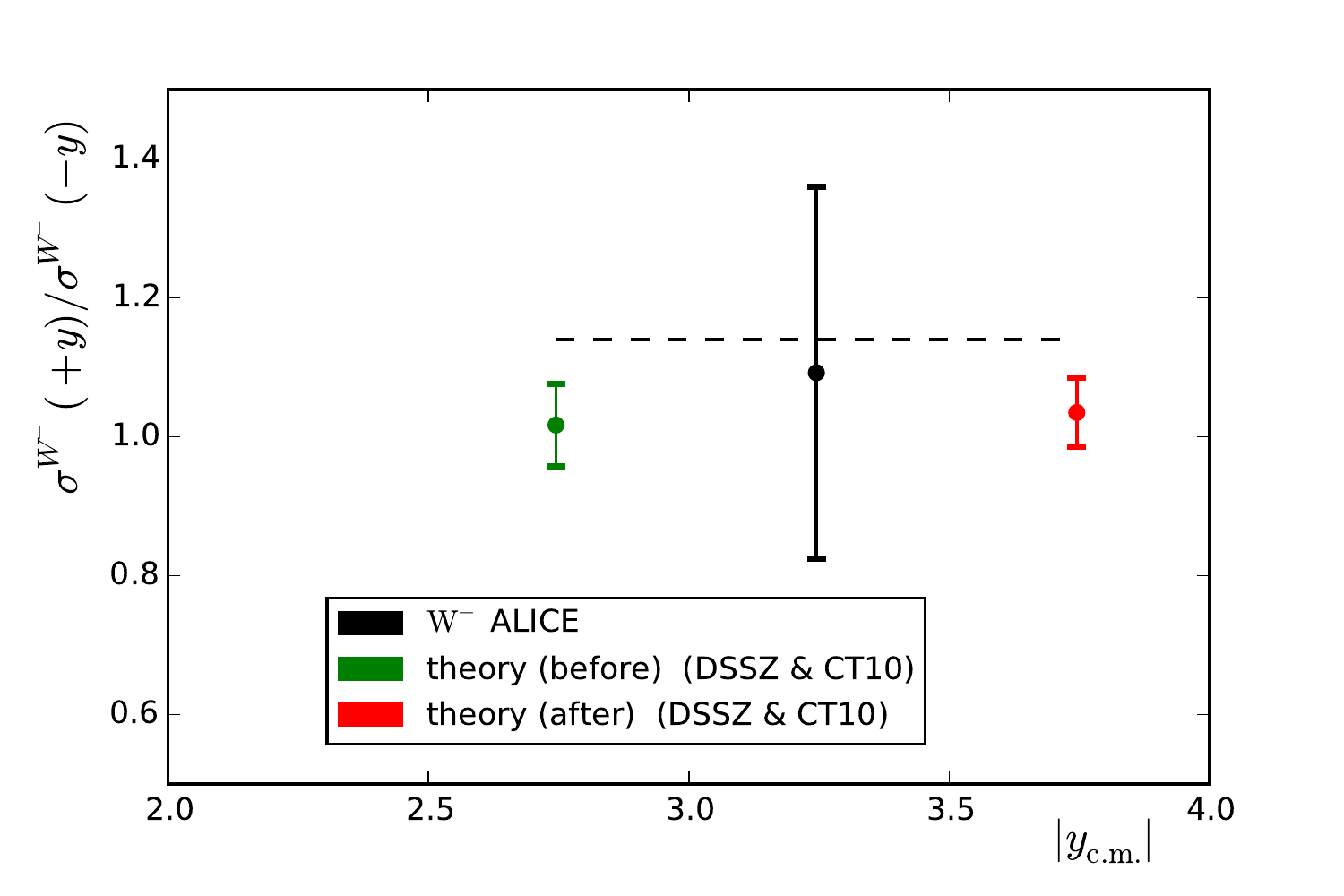}
\caption{Forward-to-backward asymmetries based on $W^{\pm}$ measurements by the ALICE collaboration. The upper (lower) graphs correspond to the theoretical calculation with EPS09 (DSSZ) nuclear PDFs. The comparisons with no nuclear effects are included as dashed lines. The rapidity values at the  horizontal axes are only indicative as the rapidity bins are different in forward and backward directions (different results are also horizontally displaced for visibility).}
\label{fig:WM_alice}
\end{figure}

The preliminary ALICE data includes events with charged leptons having $p_{\rm T}>10 \, {\rm GeV}$ at forward ($2.03 < y_{\rm c.m}< 3.53$) and backward ($-4.46 < y_{\rm c.m.} < -2.96$) rapidities in the nucleon-nucleon centre-of-mass (c.m.) frame. From these, we constructed ``forward-to-backward'' ratios as
\begin{equation}
A_{\rm F/B}
=\frac{\sigma_{W^\pm}(2.03 < y_{\rm c.m}< 3.53)}{\sigma_{W^\pm}(-4.46 < y_{\rm c.m.} < -2.96)} \, .
\label{eq:a_fb}
\end{equation}
A data-versus-theory comparison is presented in Figure~\ref{fig:WM_alice}. While the theoretical predictions do agree with the experimental values, the experimental error bars are quite large. 
Table~\ref{tab:w} (the left-hand columns) lists the corresponding values of $\chi^{2}$ before the reweighting together with those obtained assuming no nuclear modifications in PDFs. It is clear that these data have no resolution with respect to the nuclear effects in PDFs.


\begin{figure}[h!]
\centering
\includegraphics[width=0.5\textwidth]{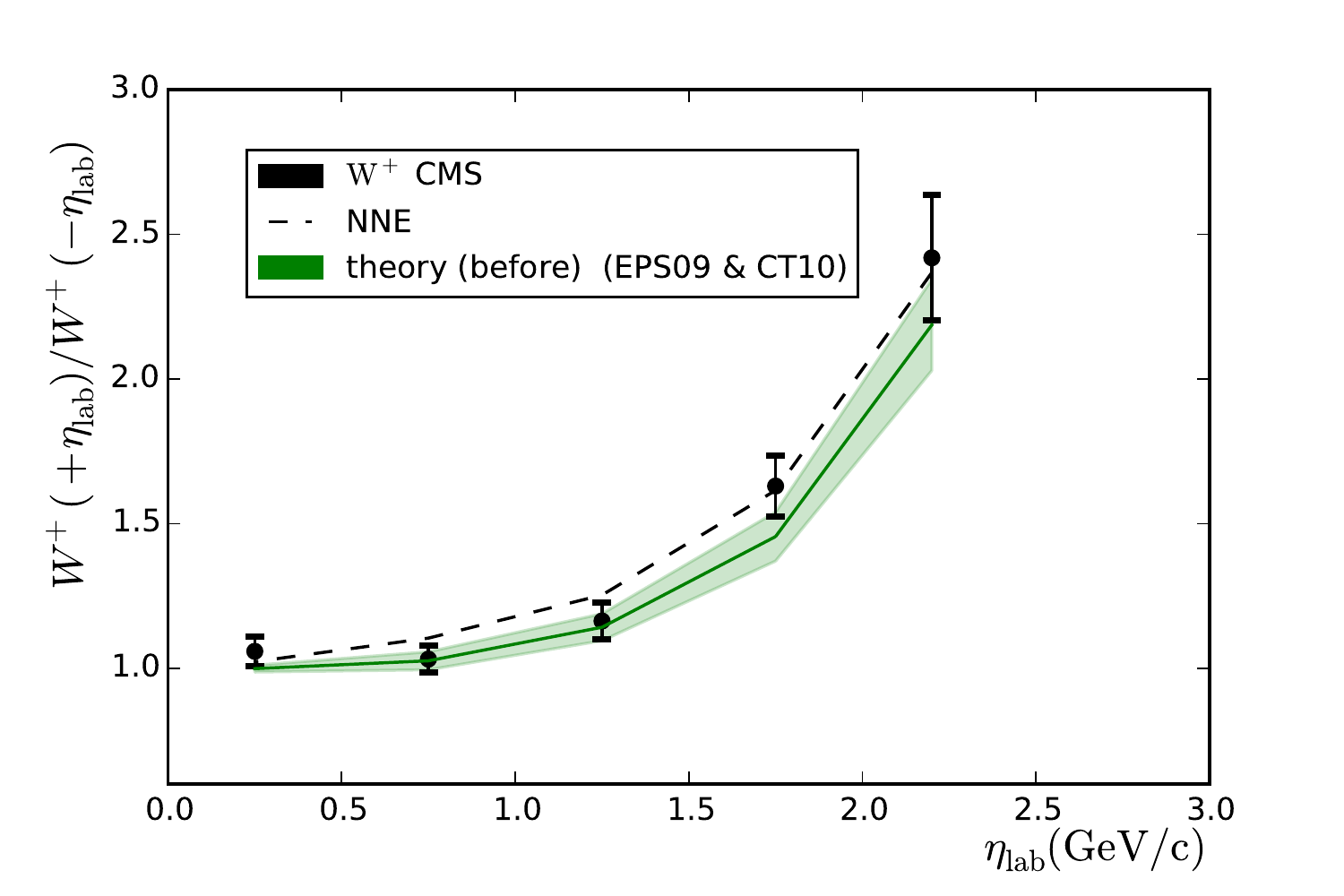} \hspace{-0.5cm}
\includegraphics[width=0.5\textwidth]{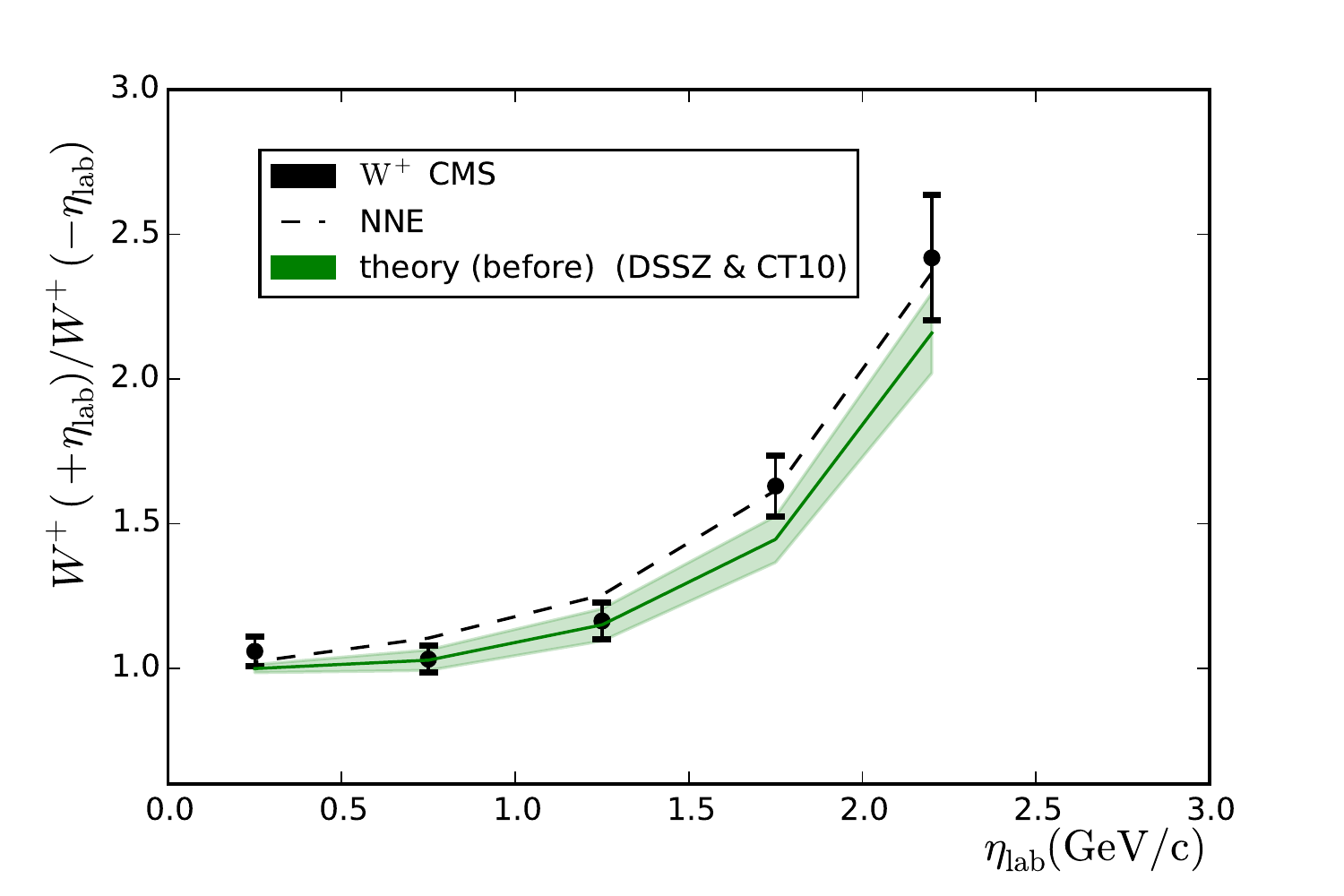}
\includegraphics[width=0.5\textwidth]{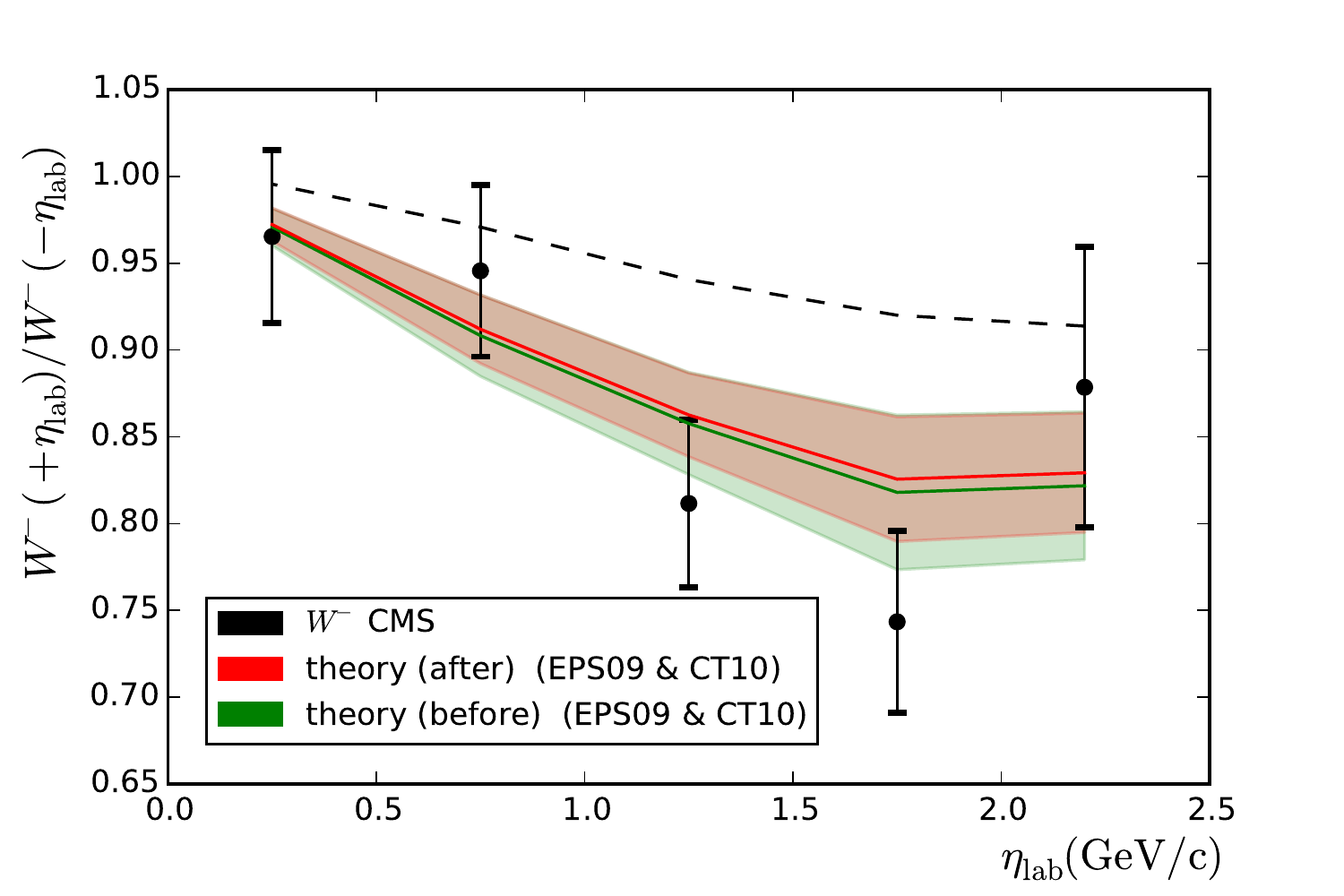} \hspace{-0.5cm}
\includegraphics[width=0.5\textwidth]{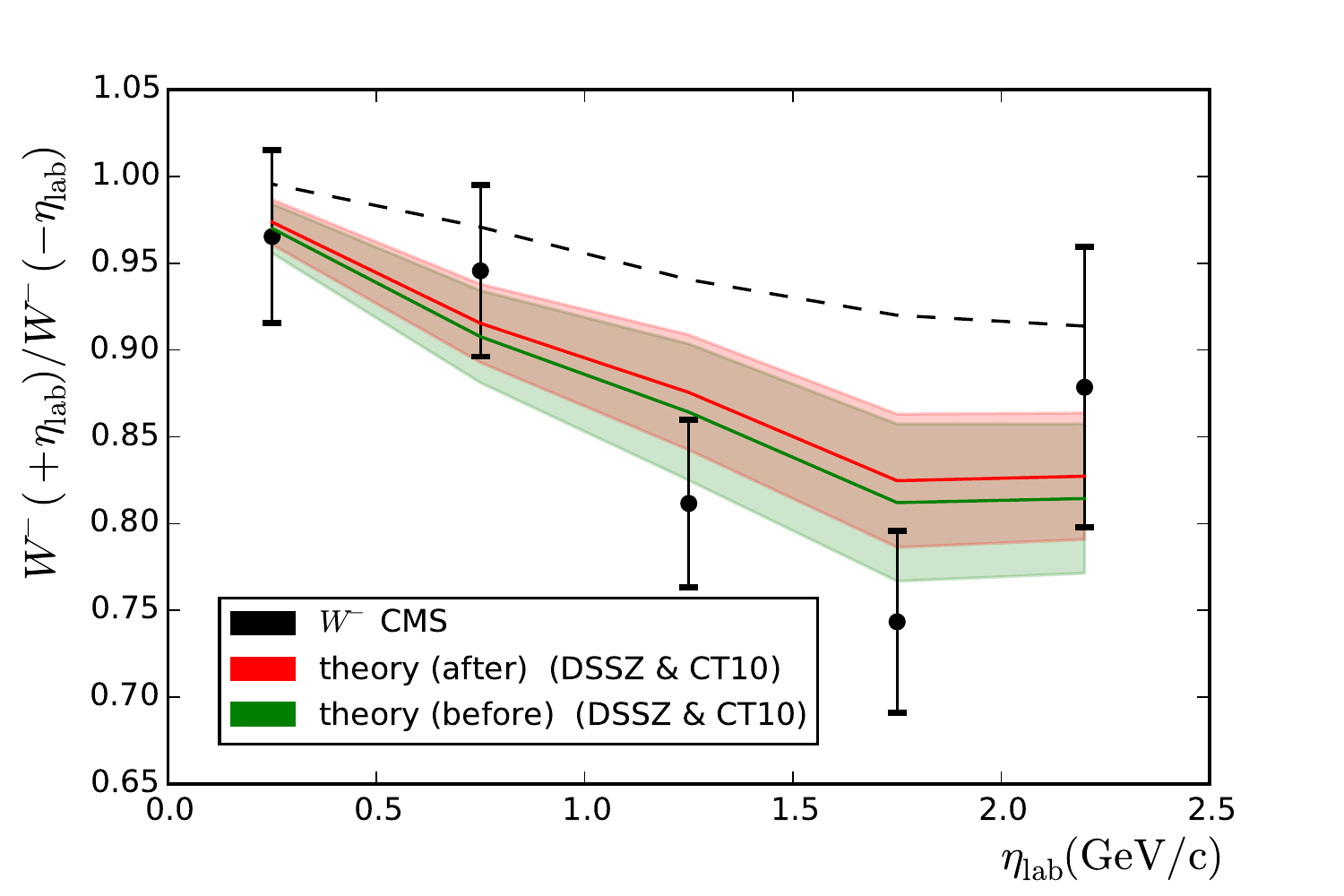}
\caption{Forward-to-backward asymmetries for $W^{+}$ (upper panels) and $W^{-}$ (lower panels) measured by the CMS collaboration \cite{Khachatryan:2015hha}, as a function of the charged-lepton pseudorapidity in the laboratory frame. The left-hand (right-hand) graphs correspond to the theoretical calculations with EPS09 (DSSZ) nPDFs. Results with no nuclear effects are included as dashed lines.}
\label{fig:WM_cms}
\end{figure}

The CMS collaboration has measured laboratory-frame pseudorapidity ($\eta_{\rm lab}$) dependent differential cross sections in the range $|\eta_{\rm lab}| < 2.4$ with the transverse momentum of the measured leptons $p_{\rm T} > 25 \, {\rm GeV}$. The measured forward-to-backward ratios are compared to the theory computations in Figure~\ref{fig:WM_cms} and the $\chi^2$ values are given in Table~\ref{tab:w} (the right-hand columns). While the $W^+$ data are roughly compatible with all the PDF combinations, the $W^-$ data show a clear preference for nuclear corrections as implemented in EPS09 and DSSZ. These measurements probe the nuclear PDFs approximately in the range $0.002 \lesssim x \lesssim 0.3$ (from most forward to most backward bin), and the nuclear effects in the forward-to-backward ratio result from the sea-quark shadowing (small $x$) becoming divided by the antishadowing in valence quarks. While the impact of these data look here somewhat limited, they may be helpful for constraining the flavour separation of nuclear modifications. However, as both EPS09 and DSSZ assume flavour-independent sea and valence quark modifications at the parametrisation scale  (i.e. the initial scale for DGLAP evolution), the present analysis cannot address to which extent this may happen.\footnote{During our analysis, an extraction of nPDFs with flavour separation was released \cite{Kovarik:2015cma}.}

\begin{table}[h!]
\begin{center}
\caption{Contribution of the $W^\pm$ data to the total $\chi^{2}$ before the reweighting. The numbers in parentheses are the amount of data points considered for each experiment.}
\label{tab:w}
\begin{tabular}{ccccc}
\\
${\rm PDF+nPDF}$ & $W^{+}_{\rm ALICE}$ (1) & $W^{-}_{\rm ALICE}$ (1) & $W^{+}_{\rm CMS}$ (5) & $W^{-}_{\rm CMS}$ (5) \\
\\
\hline
\\
CT10+DSSZ              & 0.750 & 0.082 & 5.953 & 4.140\\
CT10+EPS09            & 0.637 & 0.052 & 5.404 & 4.055 \\
CT10 only                  & 0.387 & 0.032 & 5.055 & 19.2272\\
MSTW2008+DSSZ  & 0.873 & 0.048 & 7.417 & 4.911 \\
MSTW2008+EPS09 & 0.760 & 0.035 & 6.892 & 5.205 \\
MSTW2008 only       & 0.443 & 0.054 & 4.364 & 22.869\\
\end{tabular}
\end{center}
\end{table}

\subsection{Z boson production}

The $Z$ boson production in its dilepton decay channel has been measured by three collaborations: CMS \cite{CMS:2015vqa}, ATLAS \cite{Aad:2015gta} and LHCb \cite{Aaij:2014pvu}.\footnote{The statistical uncertainties of the two LHCb data points are huge so we do not consider them here as they provide no constraining power.} As in the case of $W^{\pm}$, the theoretical values were computed using MCFM, with all scales fixed to the invariant mass of the lepton pair. 

\begin{figure}[h!]
\centering
\includegraphics[width=0.5\textwidth]{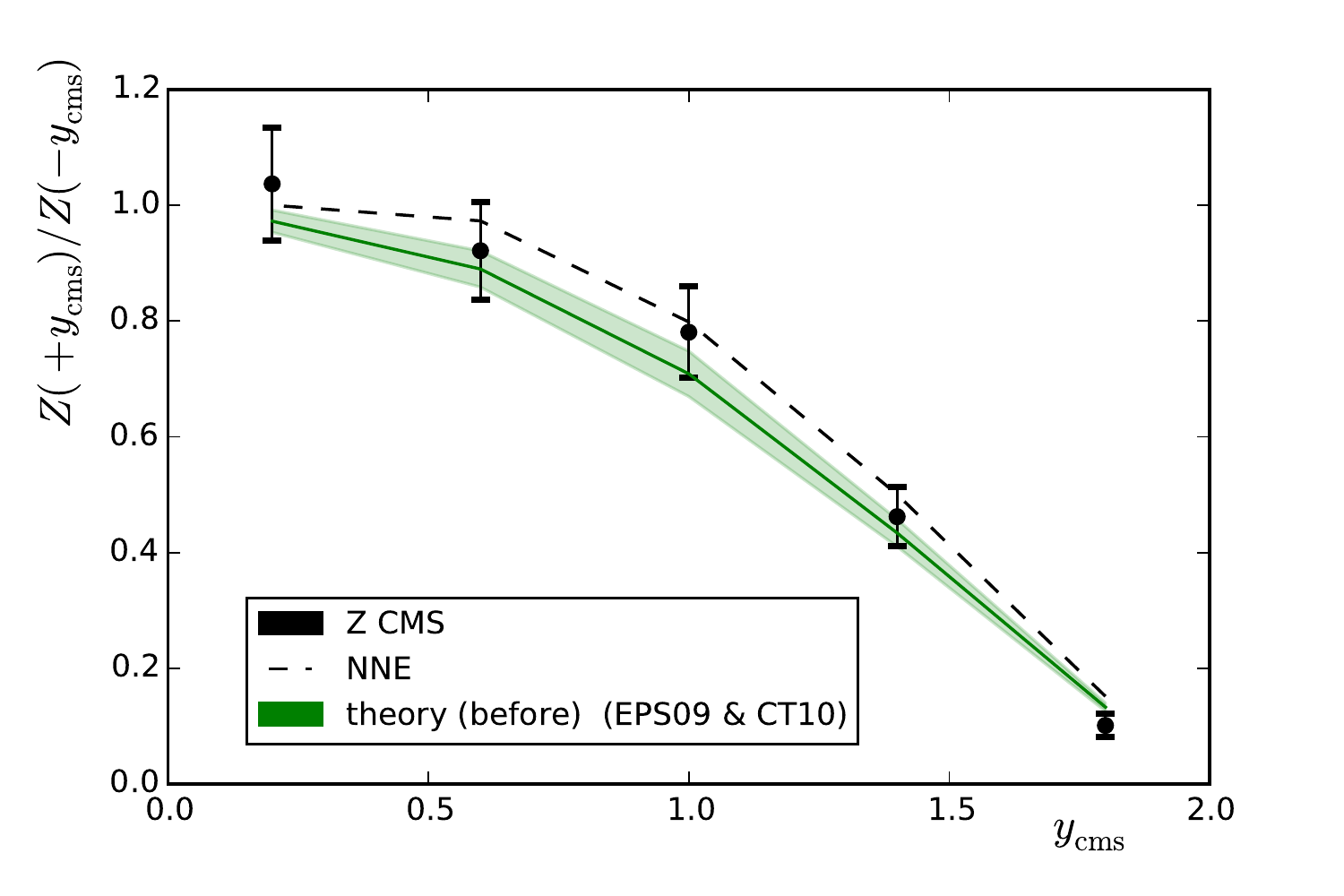} \hspace{-0.5cm} 
\includegraphics[width=0.5\textwidth]{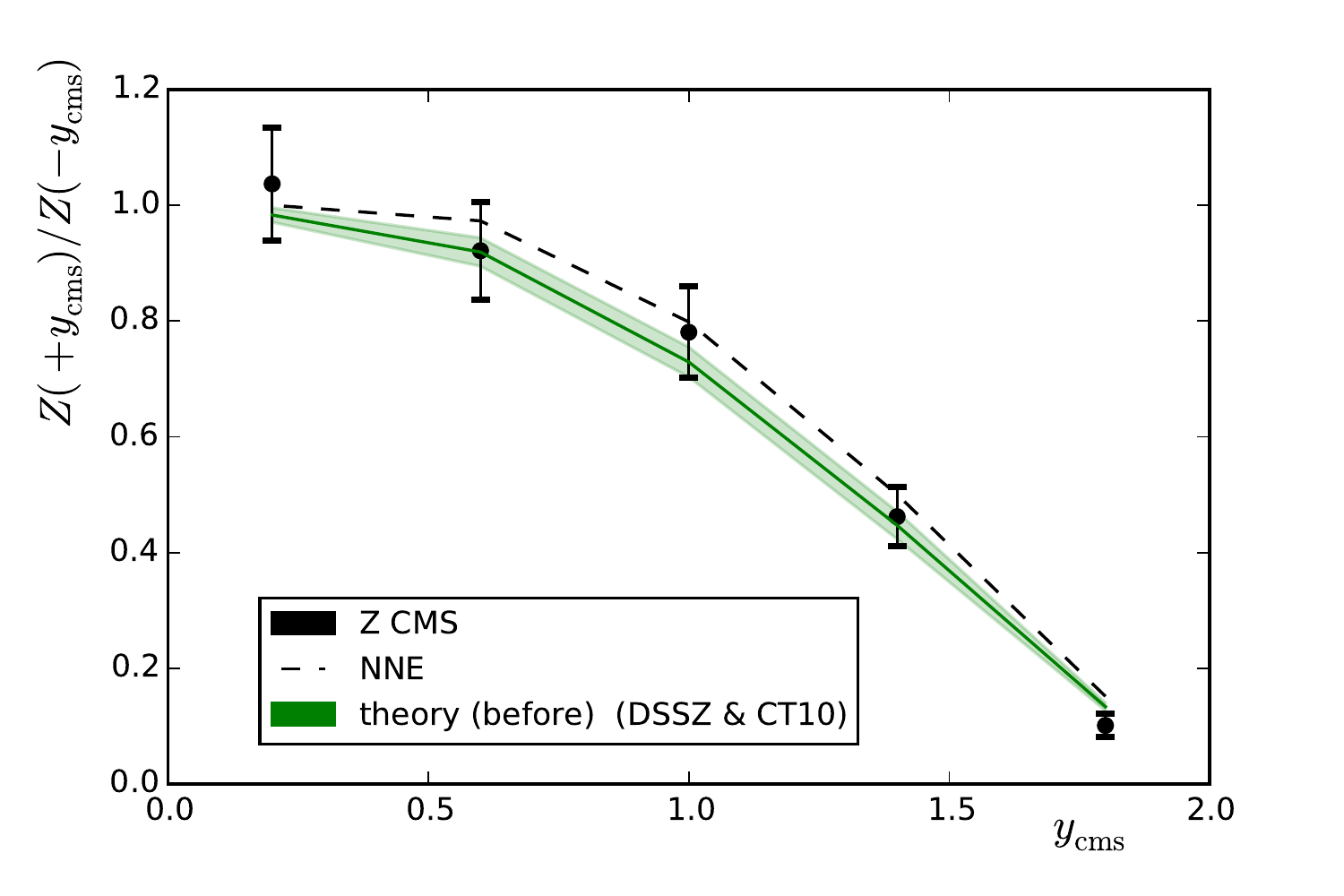} 
\caption{Forward-to-backward asymmetry of $Z$ bosons  measured by CMS \cite{CMS:2015vqa} as a function of the lepton pair rapidity. The left-hand panel (right-hand panel) shows the predictions obtained with EPS09 (DSSZ). Results with no nuclear effects (NNE) are shown as dashed lines.}
\label{fig:Z_cms}
\end{figure}

In the case of CMS, the kinematic cuts are similar to the ones applied for  $W$ bosons: the leptons are measured within $|\eta_{\rm lab}| < 2.4$ with a slightly lower minimum $p_{\rm T}$ for both leptons ($p_{\rm T} > 20$ GeV), and $60 \, {\rm GeV} < M_{l^{+}l^{-}} < 120 \, {\rm GeV}$. The $A_{\rm F/B}$ data are binned as a function of $y^{l^{+}l^{-}}_{\rm c.m.}$ (rapidity of the lepton pair). Figure~\ref{fig:Z_cms} presents a comparison between the data and theory values before the reweighting (NNE stands for no nuclear modification of parton densities but includes isospin effects) and Table~\ref{tab:z} (the right-hand column) lists the $\chi^2$ values. The data appear to slightly prefer the calculations which include nuclear modifications. Similarly to the case of $W$ production, the use of nuclear PDFs eads to a suppression in $A_{\rm F/B}$. The rapid fall-off of $A_{\rm F/B}$ towards large $y^{l^{+}l^{-}}_{\rm c.m.}$ comes from the fact that the lepton pseudorapidity acceptance is not symmetric in the nucleon-nucleon c.m. frame. Indeed the range $|\eta_{\rm lab}| < 2.4$ translates to $-2.865 < \eta_{\rm c.m.} < 1.935$ and since there is less open phase space in the forward direction, the cross sections at a given $y^{l^{+}l^{-}}_{\rm c.m.}$ tend to be lower than those at $-y^{l^{+}l^{-}}_{\rm c.m.}$. This is clearly an unwanted feature since it gives rise to higher theoretical uncertainties (which we ignore in the present study) than if a symmetric acceptance (e.g. $-1.935 < \eta_{\rm c.m.} < 1.935$) had been used.

\begin{table}[h!]
\begin{center}
\caption{
As Table~\ref{tab:w} but for Z production.
}
\label{tab:z}
\begin{tabular}{ccc}
\\
${\rm PDF+nPDF}$ & $Z_{\rm ATLAS}$ (7) & $Z_{\rm CMS}$ (5)\\
\\
\hline
\\
CT10+DSSZ              & 11.465 & 3.385 \\
CT10+EPS09            & 9.815 & 4.182 \\
CT10 only                  & 25.177 & 7.336 \\
MSTW2008+DSSZ  & 10.989 & 3.079 \\
MSTW2008+EPS09 & 9.689 & 4.193 \\
MSTW2008 only       & 24.659 & 6.834 \\
\end{tabular}
\end{center}
\end{table}

\begin{figure}[h!]
\centering
\includegraphics[width=0.5\textwidth]{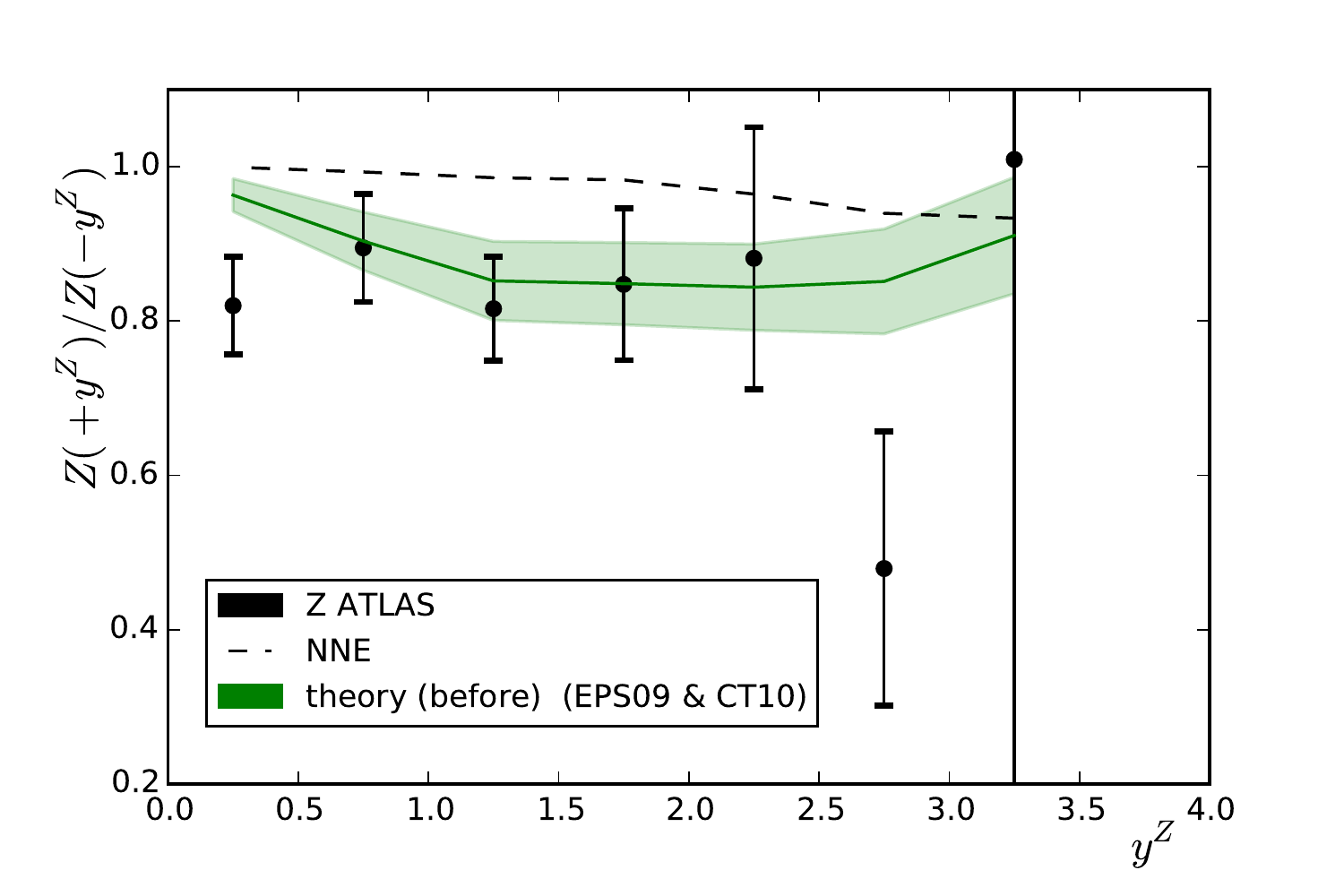} \hspace{-0.5cm}
\includegraphics[width=0.5\textwidth]{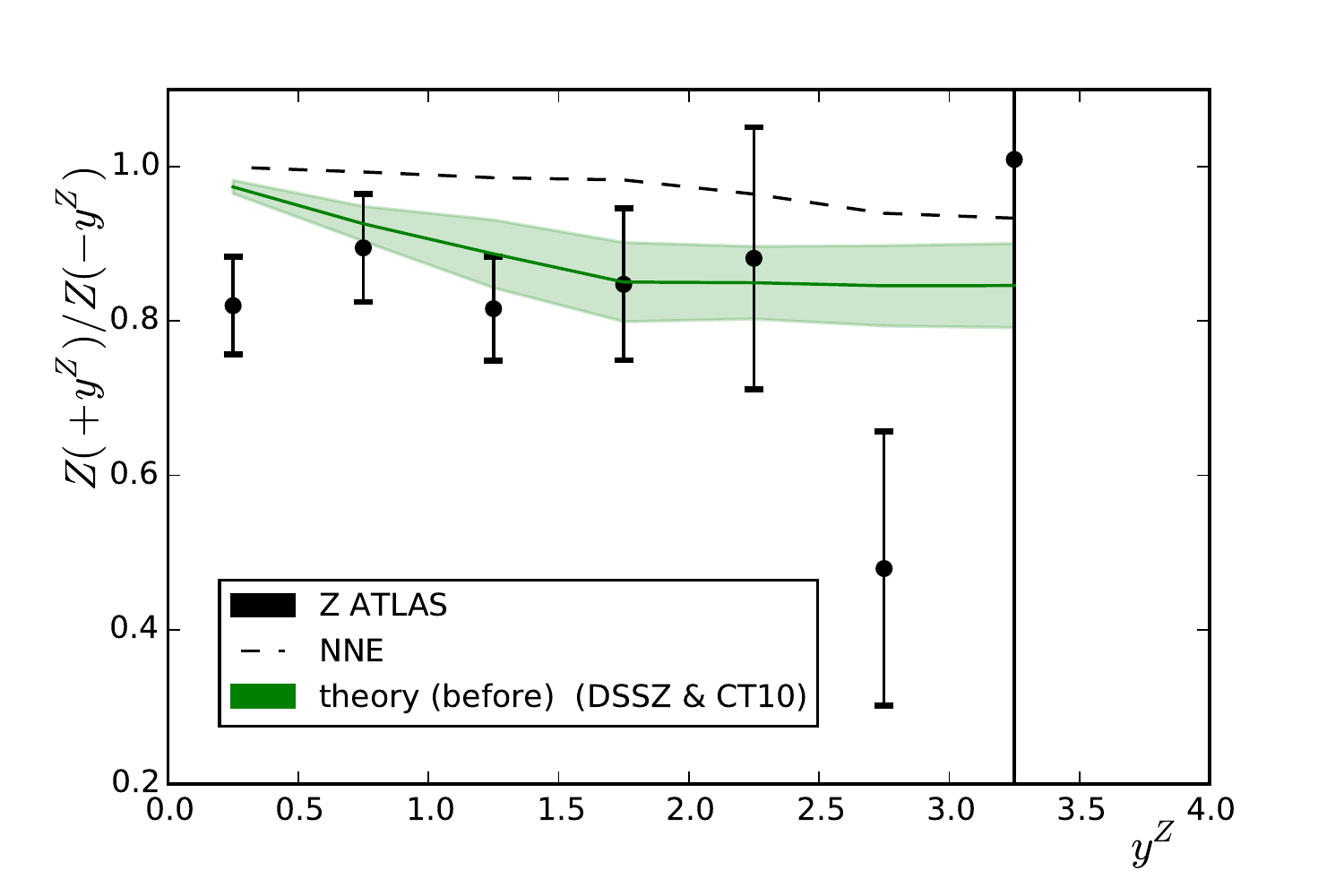} 
\caption{As in Figure~\ref{fig:Z_cms} but for the ATLAS measurement.}
\label{fig:Z_atlas}
\end{figure}

The ATLAS data correspond to the full phase space of the daughter leptons within $66 \, {\rm GeV} < M_{l^{+}l^{-}} < 116 \, {\rm GeV}$ and $|y^{Z}_{\rm c.m.}|<3.5$. The data are only availabe as absolute cross sections from which we have constructed the forward-to-backward ratio $A_{\rm F/B}$. A comparison between the theoretical predictions (with and without nuclear modifications) and the experimental values before the reweighting can be seen in Figure~\ref{fig:Z_atlas} and the $\chi^2$ values are given in Table~\ref{tab:z} (the left-hand column). The calculations including the nuclear modifications are now clearly preferred. For the larger phase space, $A_{\rm F/B}$ is now significantly closer to unity than in Figure~\ref{fig:Z_cms}.

\subsection{Jets \& di-jets}

Jet and di-jet distributions were computed at NLO \cite{Frixione:1995ms,Frixione:1997np,Frixione:1997ks} and compared with the results from the ATLAS \cite{ATLAS:2014cpa} and CMS \cite{Chatrchyan:2014hqa} collaborations, respectively. The factorisation and renormalisation scales were fixed to half the sum of the transverse energy of all 2 or 3 jets in the event. 
For ATLAS jets we used the anti-$k_T$ algorithm \cite{Cacciari:2008gp} with $R=0.4$. For the CMS di-jets we used the anti-$k_T$ algorithm with $R=0.3$ and only jets within the acceptance $|\eta_{\rm jet}|<3$ were accepted, and the hardest (1) and next-to-hardest (2) jet within the acceptance had to fulfil the conditions $p_{T{\rm jet},1}>120$ GeV/c,  $p_{T{\rm jet},2}>30$ GeV/c and their azimuthal distance $\Delta \phi_{12}>2\pi /3$.

\begin{figure}
\centering
\includegraphics[width=.5\linewidth]{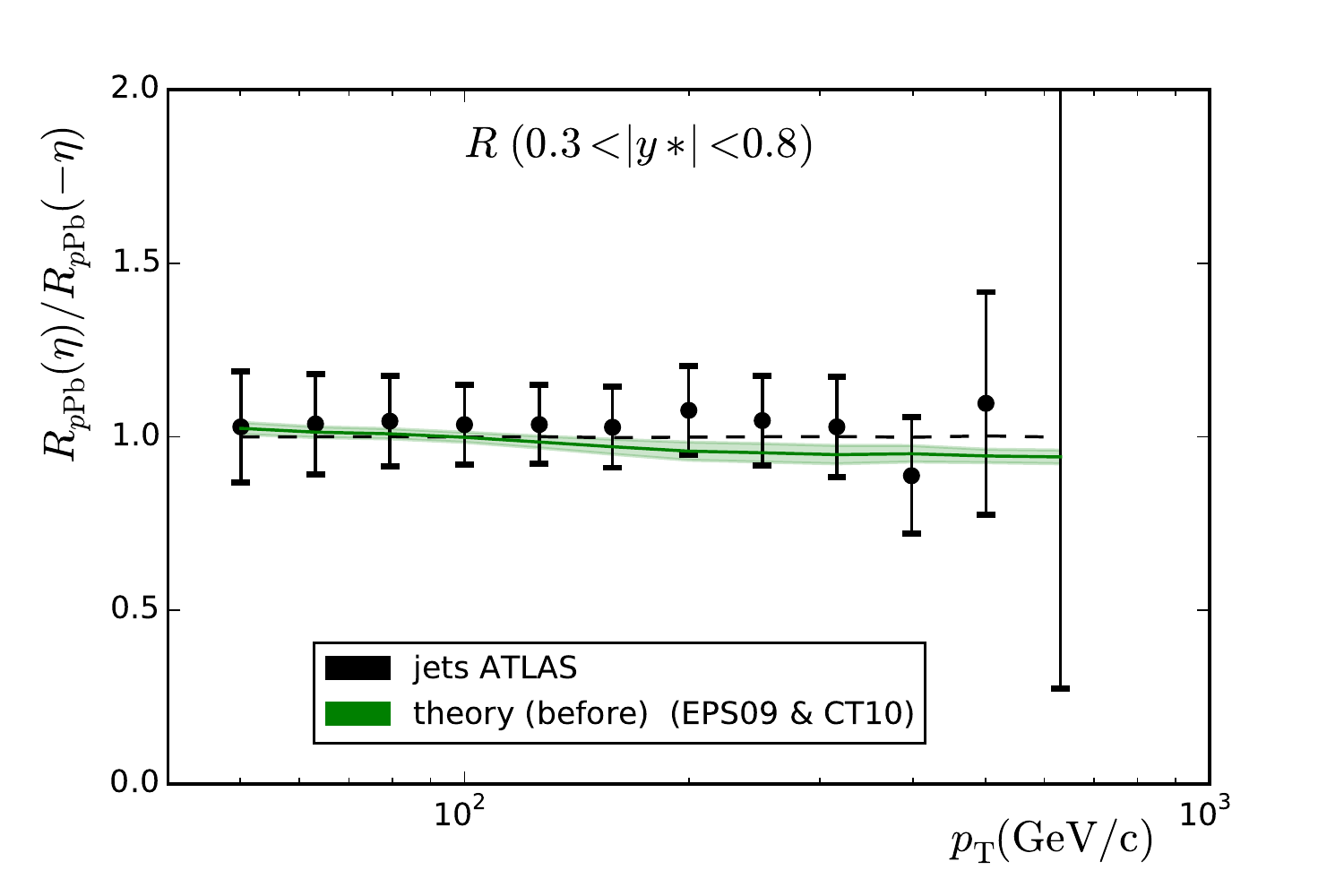} \hspace{-0.5cm}
\includegraphics[width=.5\linewidth]{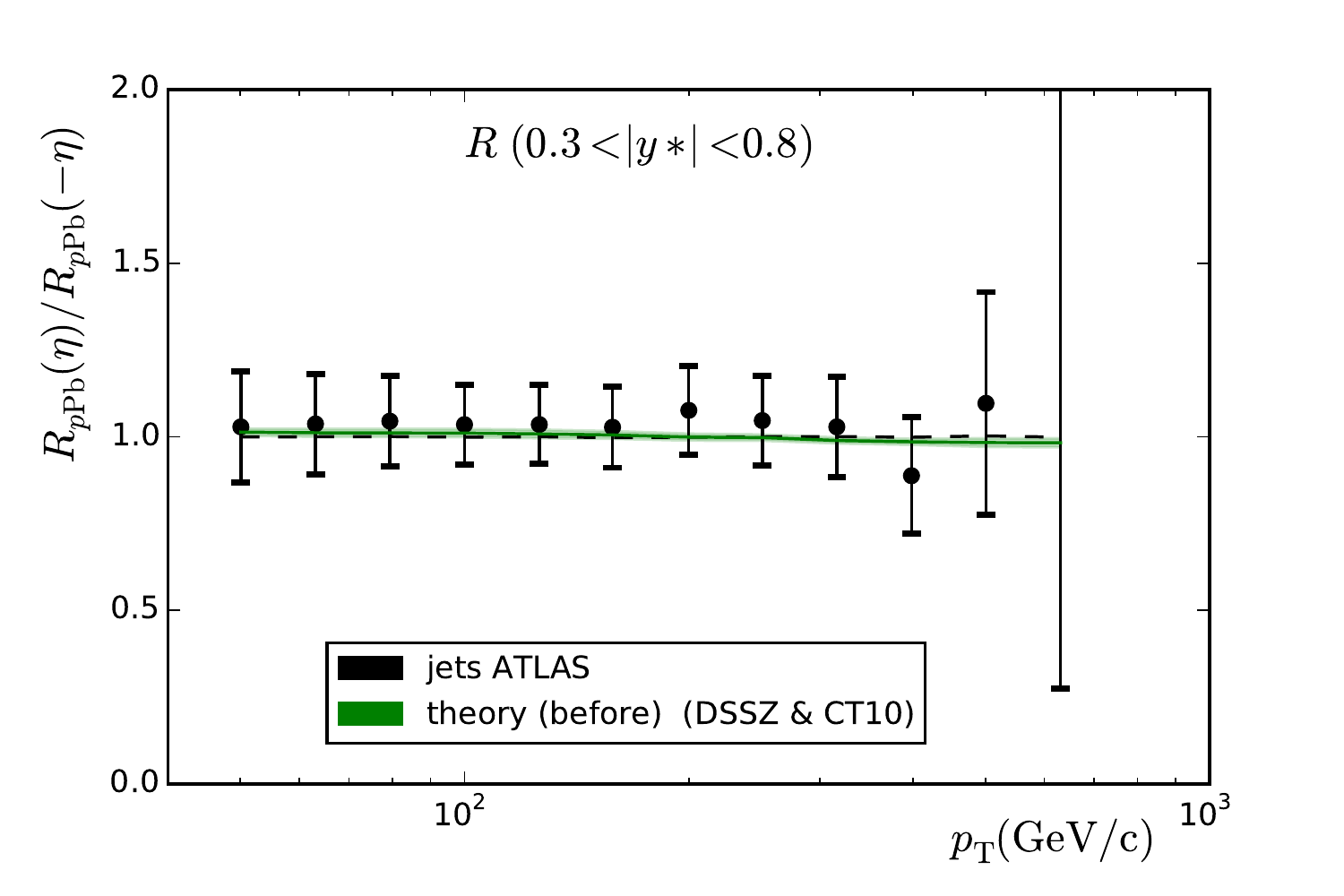}
\includegraphics[width=.5\linewidth]{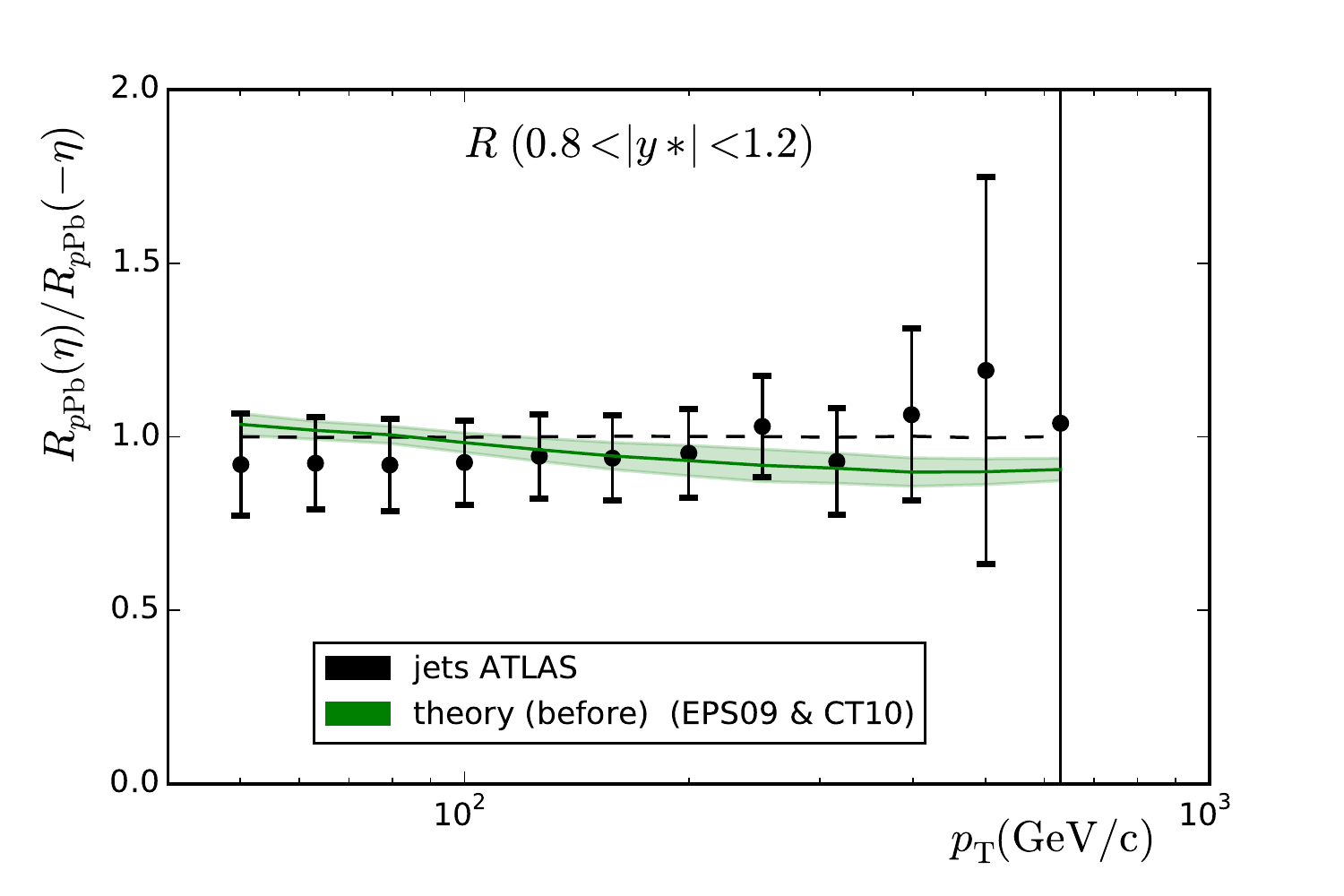} \hspace{-0.5cm}
\includegraphics[width=.5\linewidth]{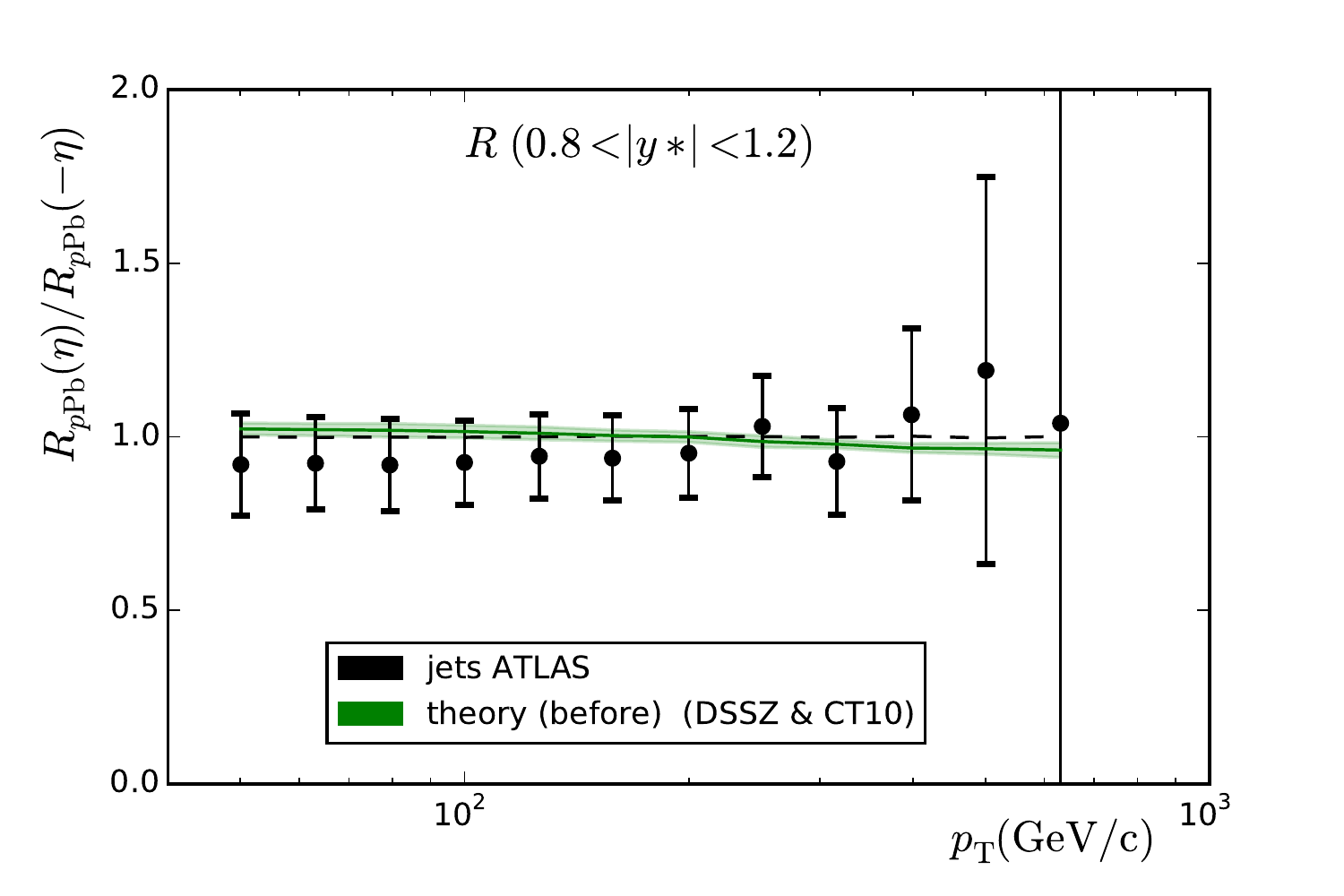}
\includegraphics[width=.5\linewidth]{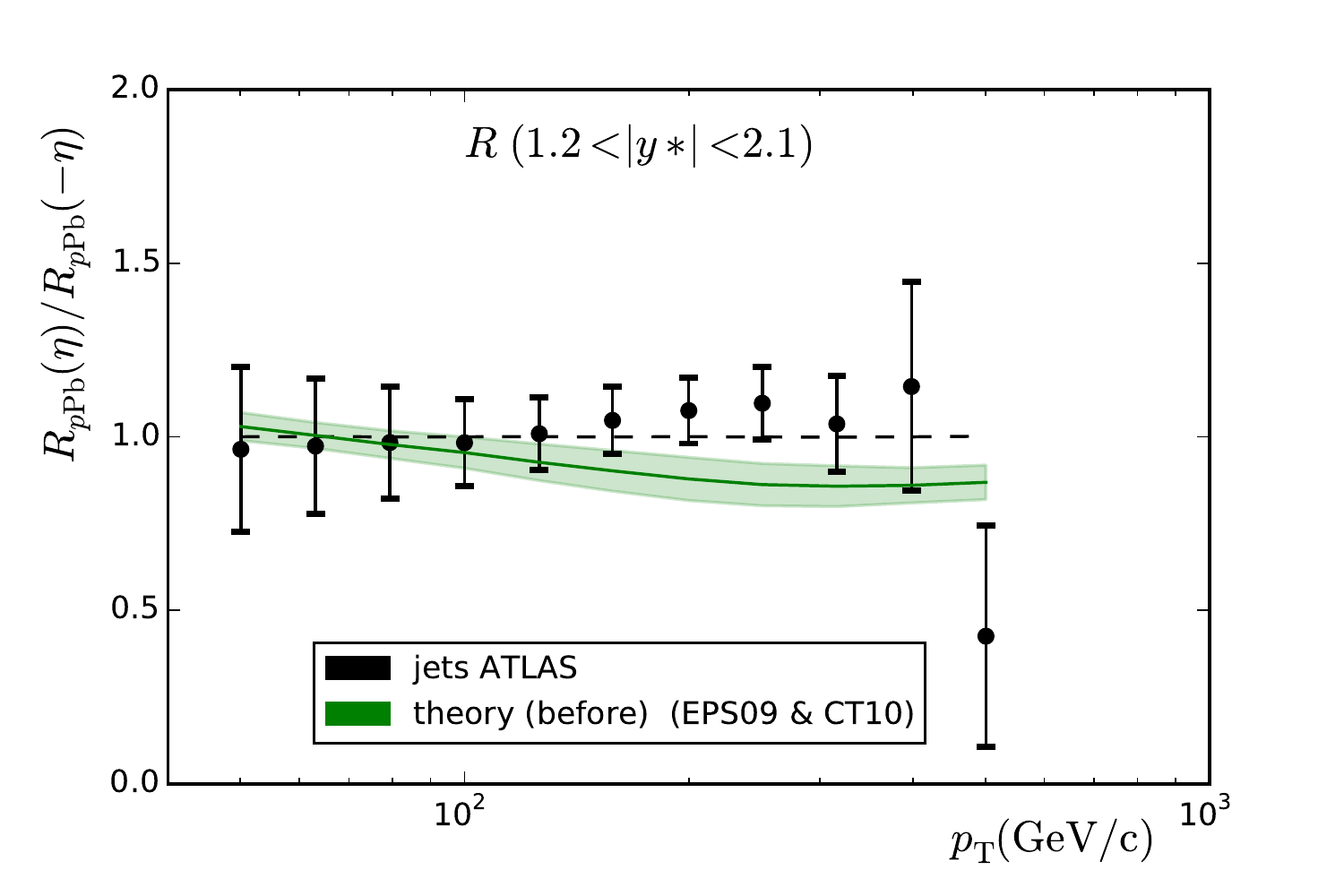} \hspace{-0.5cm}
\includegraphics[width=.5\linewidth]{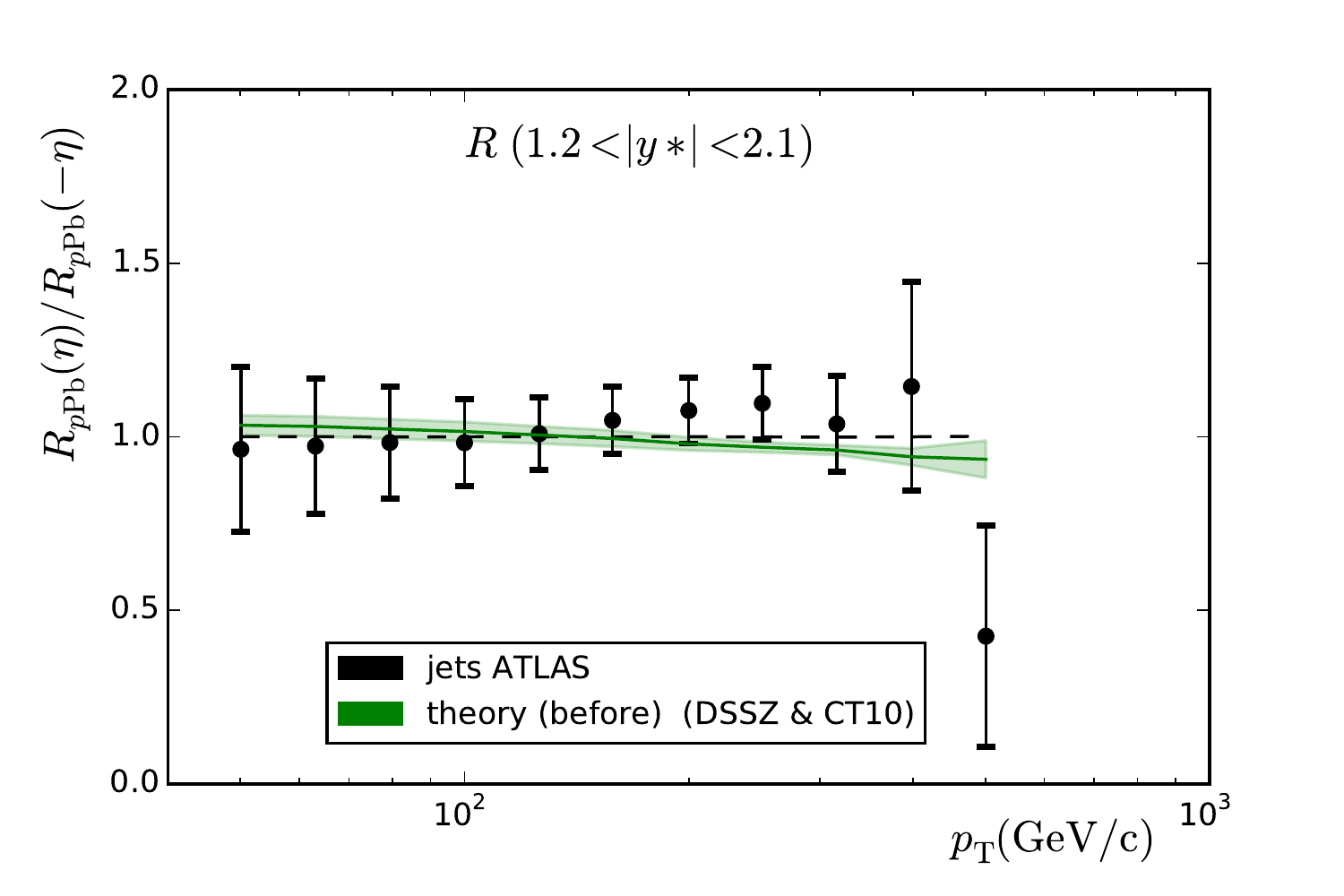}
\caption{Forward-to-backward ratios based on the ATLAS jet cross-section measurements as a function of jet $p_{\rm T}$. The theoretical predictions and uncertainty bands were computed using the eigenvectors of EPS09 (left) and DSSZ (right). Upper panels: $0.3 < |y*| < 0.8$. Middle panels: $0.8 < |y*| < 1.2$. Lower panels: $1.2 < |y*| < 2.1$. 
}
\label{fig:jets_atlas}
\end{figure}

The ATLAS collaboration measured jets with transverse momentum up to $1 \, {\rm TeV}$ in eight rapidity bins. Strictly speaking, these data are not minimum bias as they comprise the events within the $0-90\%$ centrality class. It is therefore somewhat hazardous to include them into the present analysis but, for curiosity, we do so anyway. The ATLAS data are available as absolute yields from which we have constructed the forward-to-backward asymmetries adding all the uncertainties in quadrature. Let us remark that, by proceeding this way, we lose the most forward $2.1 < y^* < 2.8$ and central $-0.3 < y^* < 0.3$ bins. The results before the reweighting are presented in Figure~\ref{fig:jets_atlas} and Table~\ref{tab:jdj} (left-hand column). For EPS09 the forward-to-backward ratio tends to stay below unity since at positive rapidities the spectrum gets suppressed (gluon shadowing) and enhanced at negative rapidities (gluon antishadowing). For DSSZ, the effects are milder. The data does not appear to show any systematic tendency from one rapidity bin to another which could be due to the centrality trigger imposed. Indeed, the best $\chi^2$ is achieved with no nuclear effects at all, but all values of $\chi^2/N_{\rm data}$ are very low. This is  probably due to overestimating the systematic uncertainties  by adding all errors in quadrature. It is worth mentioning here that, on the contrary to the ATLAS data, the preliminary CMS inclusive jet data \cite{CMS:2014qca} (involving no centrality selection) do show a consistent behaviour with EPS09. 


\begin{table}[h!]
\begin{center}
\caption{
As Table~\ref{tab:w} but for jets and dijets.
}
\label{tab:jdj}
\begin{tabular}{ccc}
\\
${\rm PDF+nPDF}$ & ${\rm jets_{ATLAS}}$ (35) & ${\rm dijets_{CMS}}$ (15) \\
\\
\hline
\\
CT10+DSSZ              & 11.518 & 94.441 \\
CT10+EPS09            & 23.322 & 10.526 \\
CT10 only                  & 9.785 & 116.187 \\
MSTW2008+DSSZ  & 11.629 & 56.365 \\
MSTW2008+EPS09 & 22.833 & 5.522 \\
MSTW2008 only       & 9.811 & 67.763 \\
\end{tabular}
\end{center}
\end{table}

\begin{figure}
\centering
\includegraphics[width=0.5\textwidth]{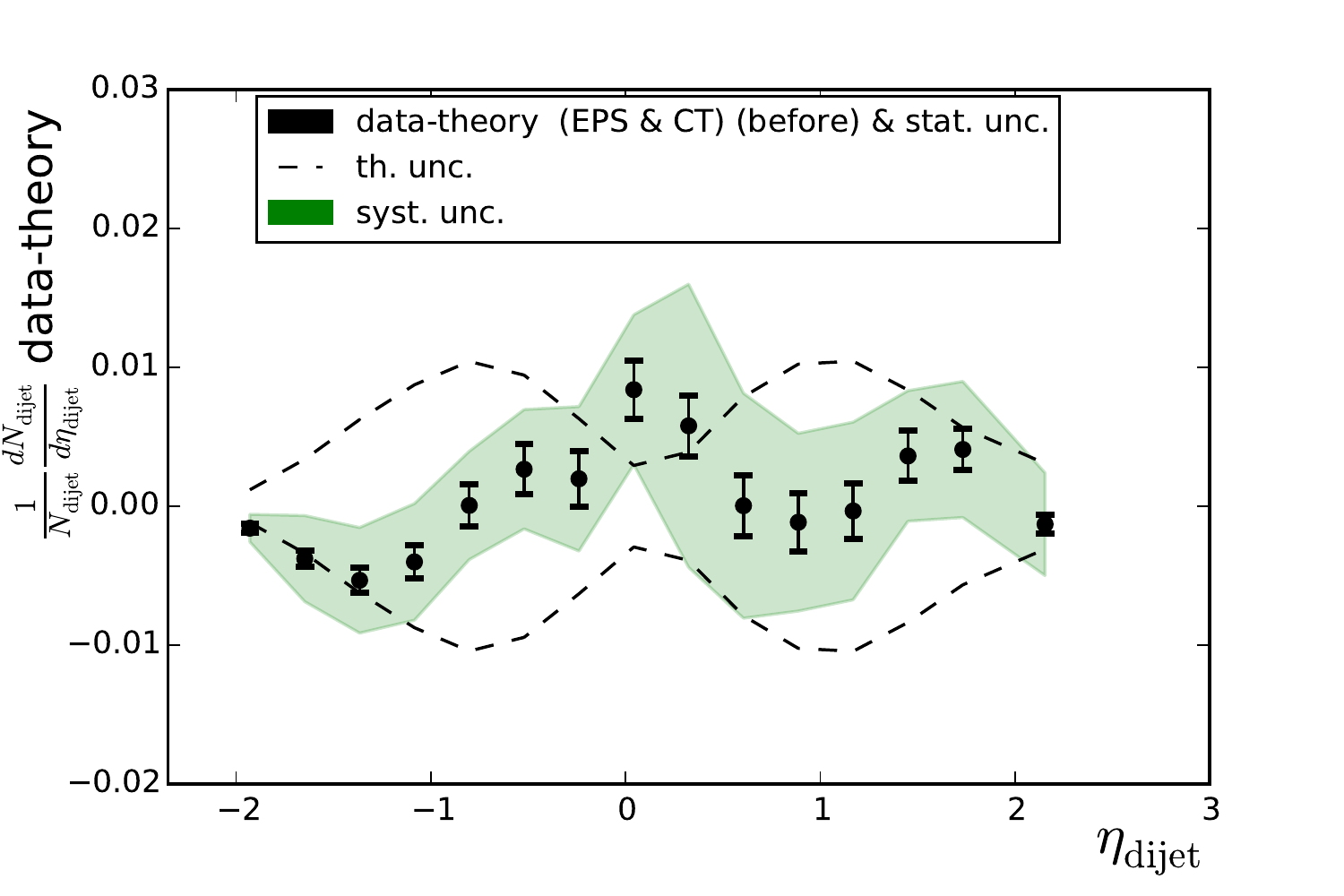} \hspace{-0.5cm}
\includegraphics[width=0.5\textwidth]{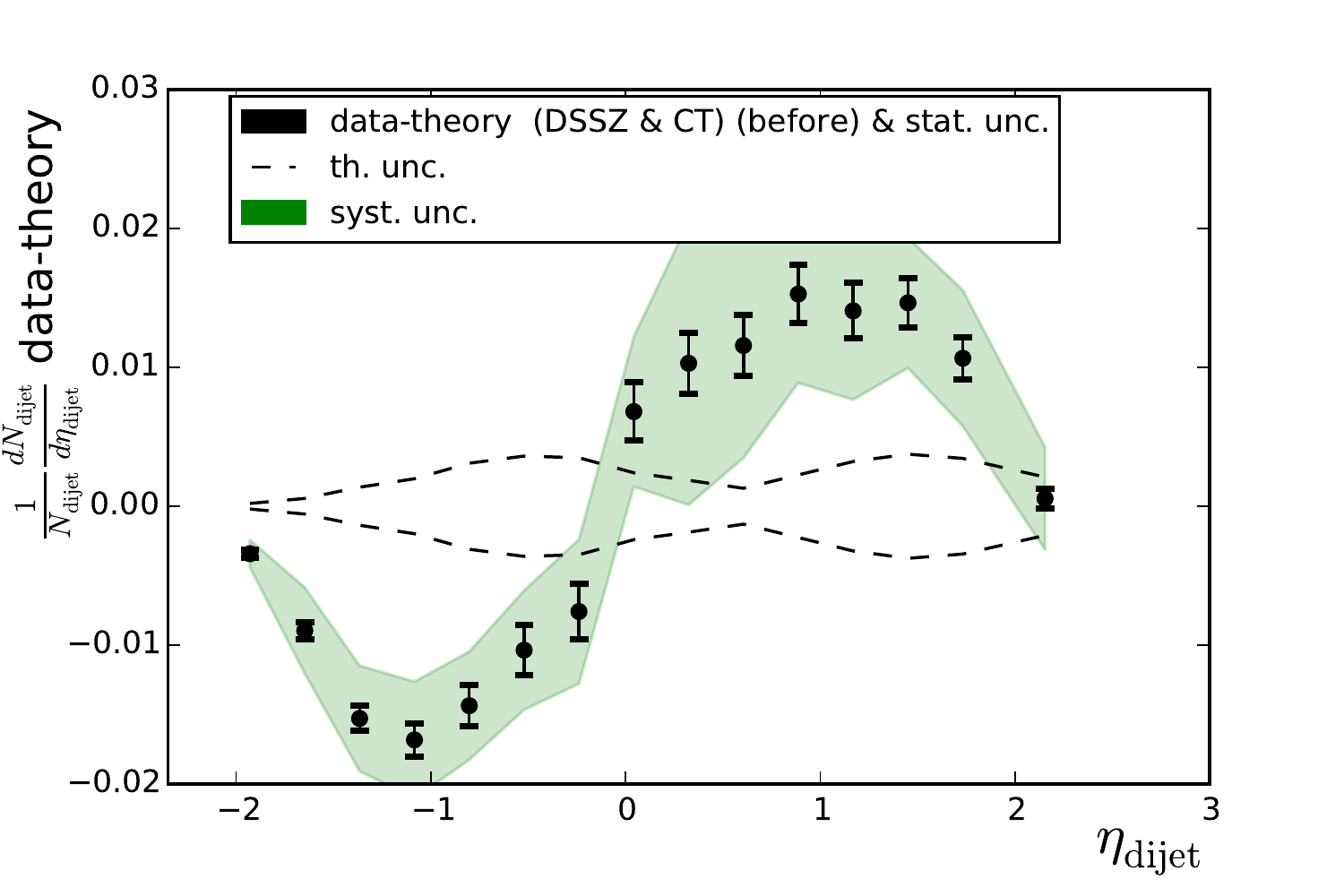}
\caption{The CMS dijet data presented as differences between the data and the theory calculations. The dashed lines correspond to the the nPDF uncertainty.
}
\label{fig:di-jets_cms}
\end{figure}

Di-jet production by the CMS collaboration \cite{Chatrchyan:2014hqa} was the subject of study in \cite{Eskola:2013aya}, where  sizeable mutual deviations between different nuclear PDFs were found. The experimental observable in this case is normalised to the total number of di-jets and the proton reference uncertainties tend to cancel to some extent, especially around midrapidity. A better cancellation would presumably be attained by considering the forward-to-backward ratios, but this would again involve the issue of correlated systematic uncertainties mentioned earlier. Comparisons between the data and theoretical predictions are shown in Figure~\ref{fig:di-jets_cms} and the $\chi^2$ values are tabulated in Table~\ref{tab:jdj} (right-hand column). The data clearly favour the use of EPS09 nPDFs, and in all other cases $\chi^2/N_{\rm data}= 3.8\ldots 7.8$ which is a clear signal of incompatibility. The better agreement follows from the gluon antishadowing and EMC effect at large $x$ present in EPS09 but not in DSSZ. However, the significant dependence of the employed free-proton PDFs is a bit alarming: indeed, one observes around 50\% difference when switching from CT10 to MSTW2008. This indicates that the cancellation of proton PDF uncertainties is not complete at all and that they must be accounted for (unlike we do here) if this observable is to be used as an nPDF constraint. The proton-proton reference data taken in Run II may improve the situation.

\subsection{Charged-particle production}

Now let us move to the analysis of charged-particle production. Here we consider both charged-hadron (ALICE \cite{Abelev:2014dsa} and CMS \cite{CMS:2013cka}) and pion (ALICE \cite{Richer:2015vqa}) production. Apart from the PDFs, the particle production depends on the fragmentation functions (FFs) which are not well constrained. Indeed, it has been shown that any of the current FFs cannot give a proper description of the experimental results \cite{d'Enterria:2013vba} on charged-hadron production. In the same reference, a kinematic cut $p_{\rm T} > 10 \, {\rm GeV}$ was advocated to avoid contaminations from other than independent parton-to-hadron ragmentation mechanism described by FFs. The same cut is applied here. Regarding the final state pions, we relaxed the requirement to $p_{\rm T} > 2 \, {\rm GeV}$, since cuts like this have been used in the EPS09 and DSSZ analyses. The theoretical values were determined with the same code as in \cite{Sassot:2010bh}, using the fragmentation functions from DSS \cite{deFlorian:2007hc} for the charged hadrons. In the case of the DSSZ nPDFs medium-modified fragmentation functions were used \cite{Sassot:2009sh}, in accordance with the way in which the RHIC pion data \cite{Adler:2006wg} were treated in the original DSSZ extraction. This is, however, not possible in the case of unidentified charged hadrons, as medium modified fragmentation functions are available for pions and kaons only.

\begin{figure}
\centering
\includegraphics[width=0.5\textwidth]{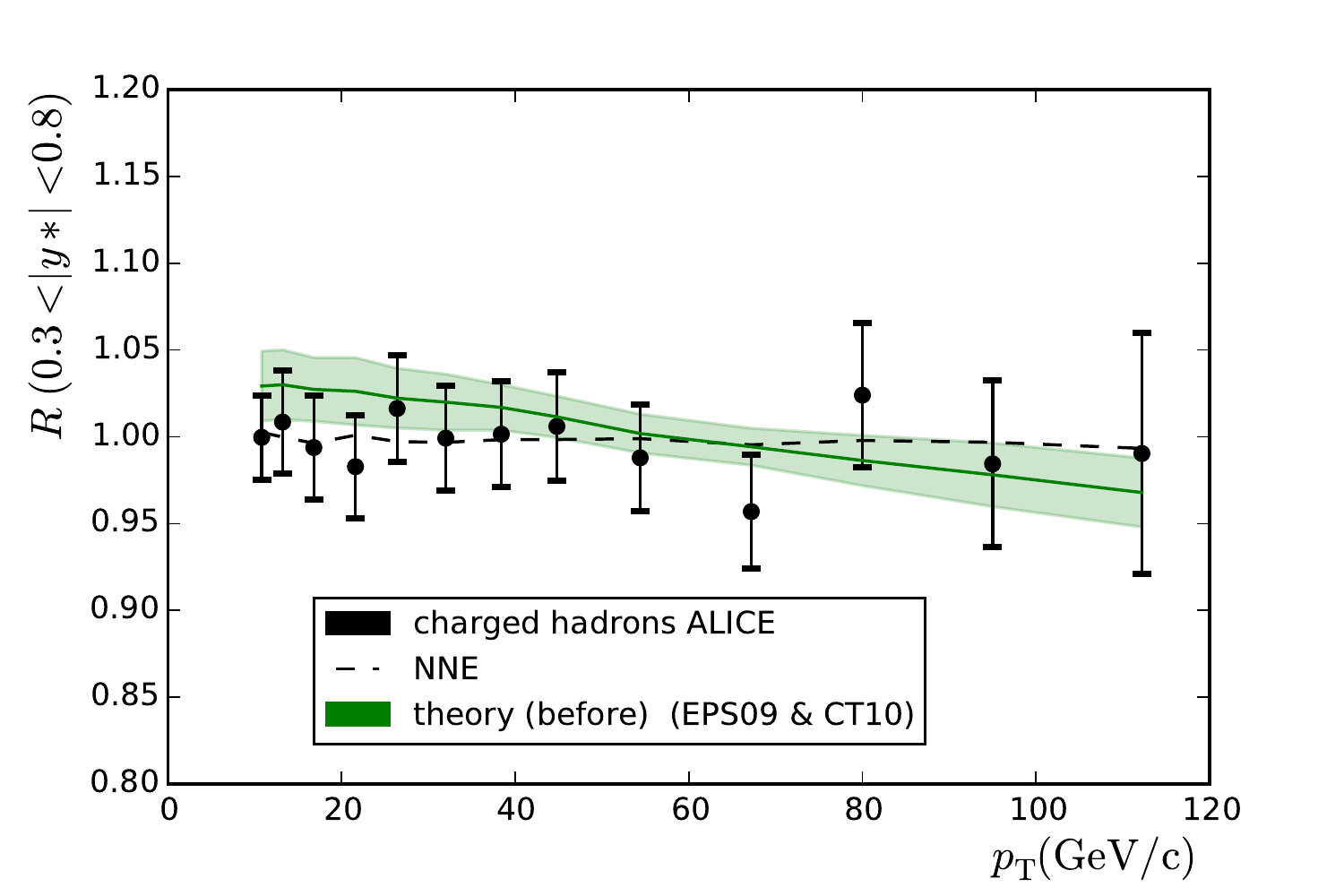} \hspace{-0.5cm}
\includegraphics[width=0.5\textwidth]{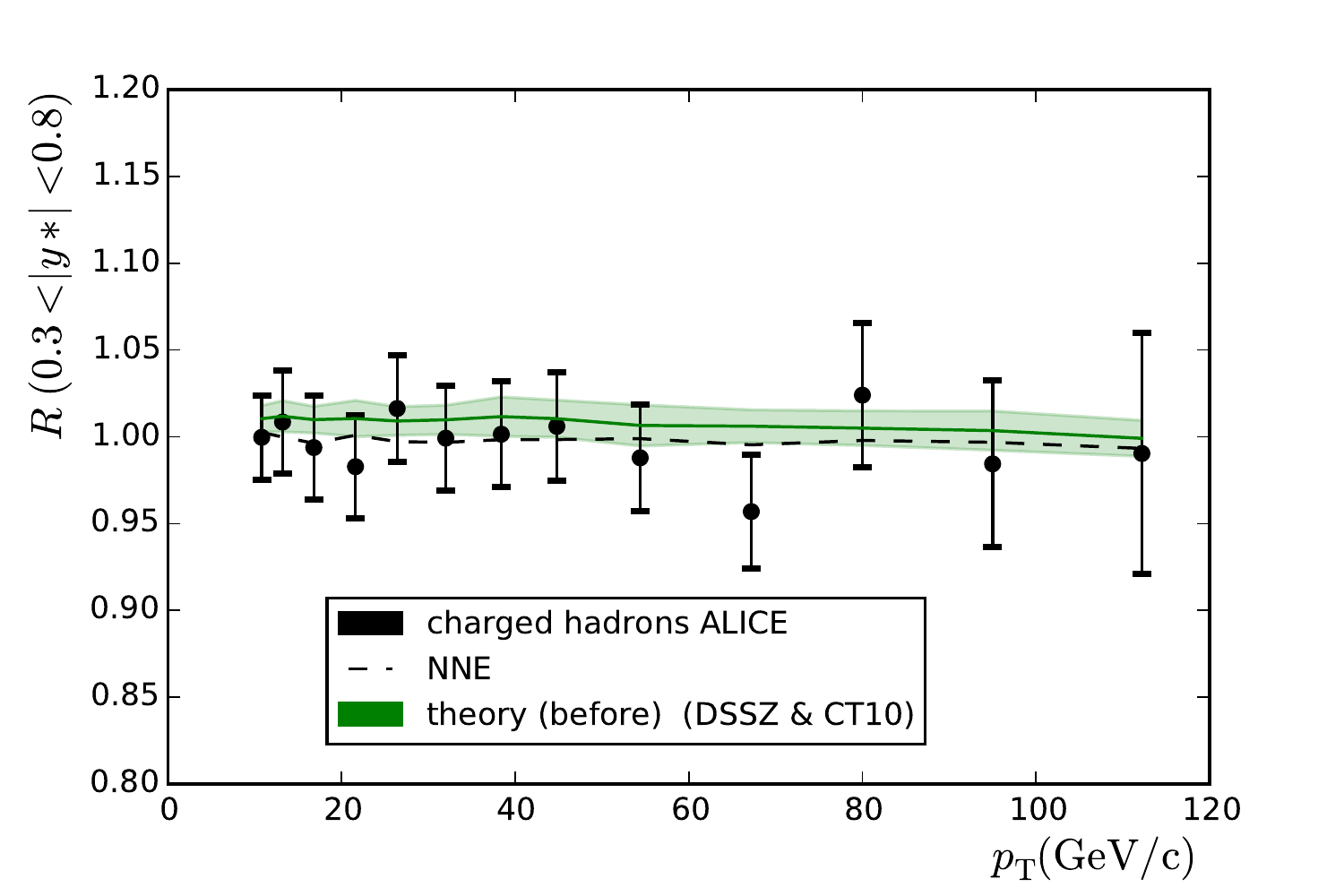}
\includegraphics[width=0.5\textwidth]{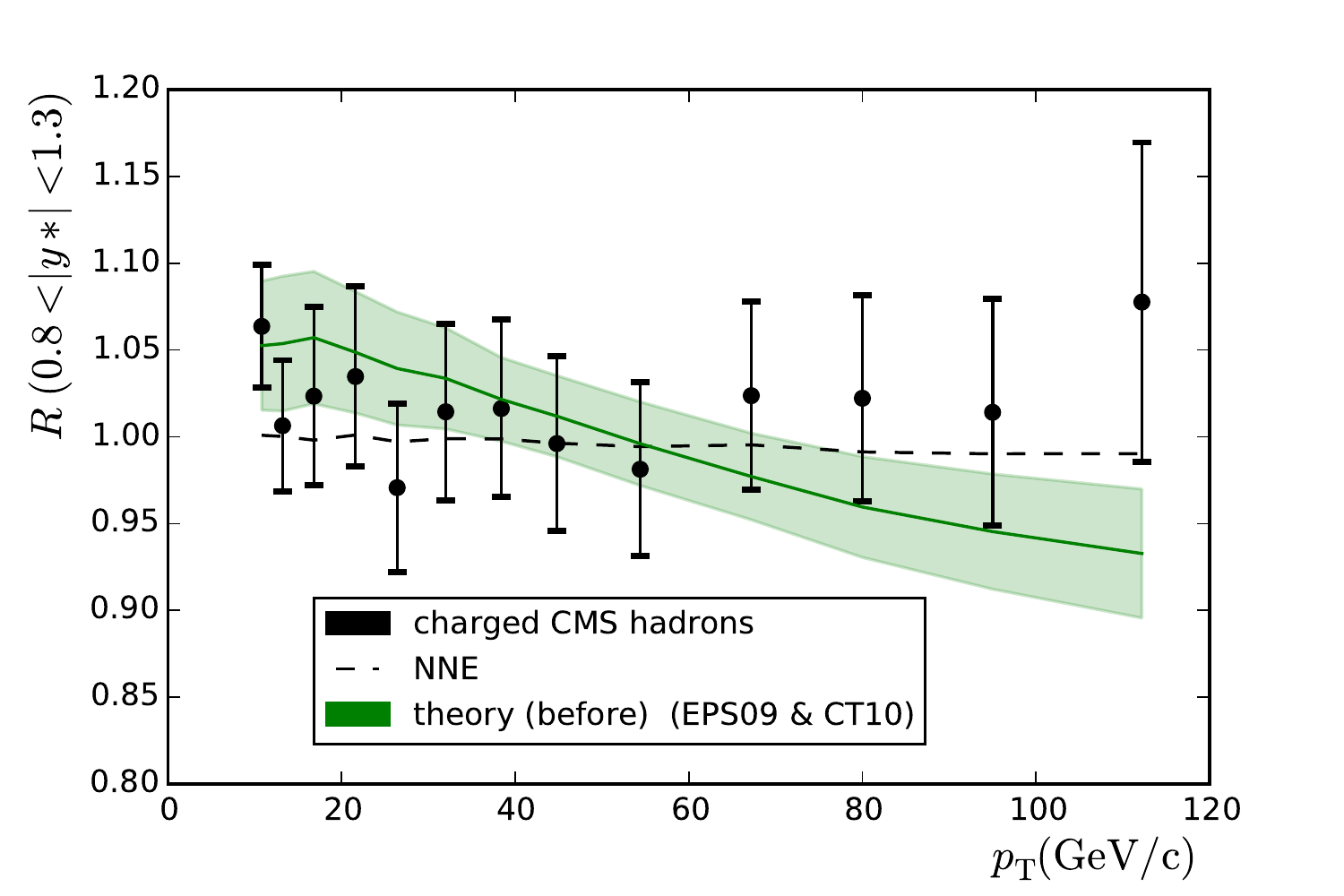} \hspace{-0.5cm}
\includegraphics[width=0.5\textwidth]{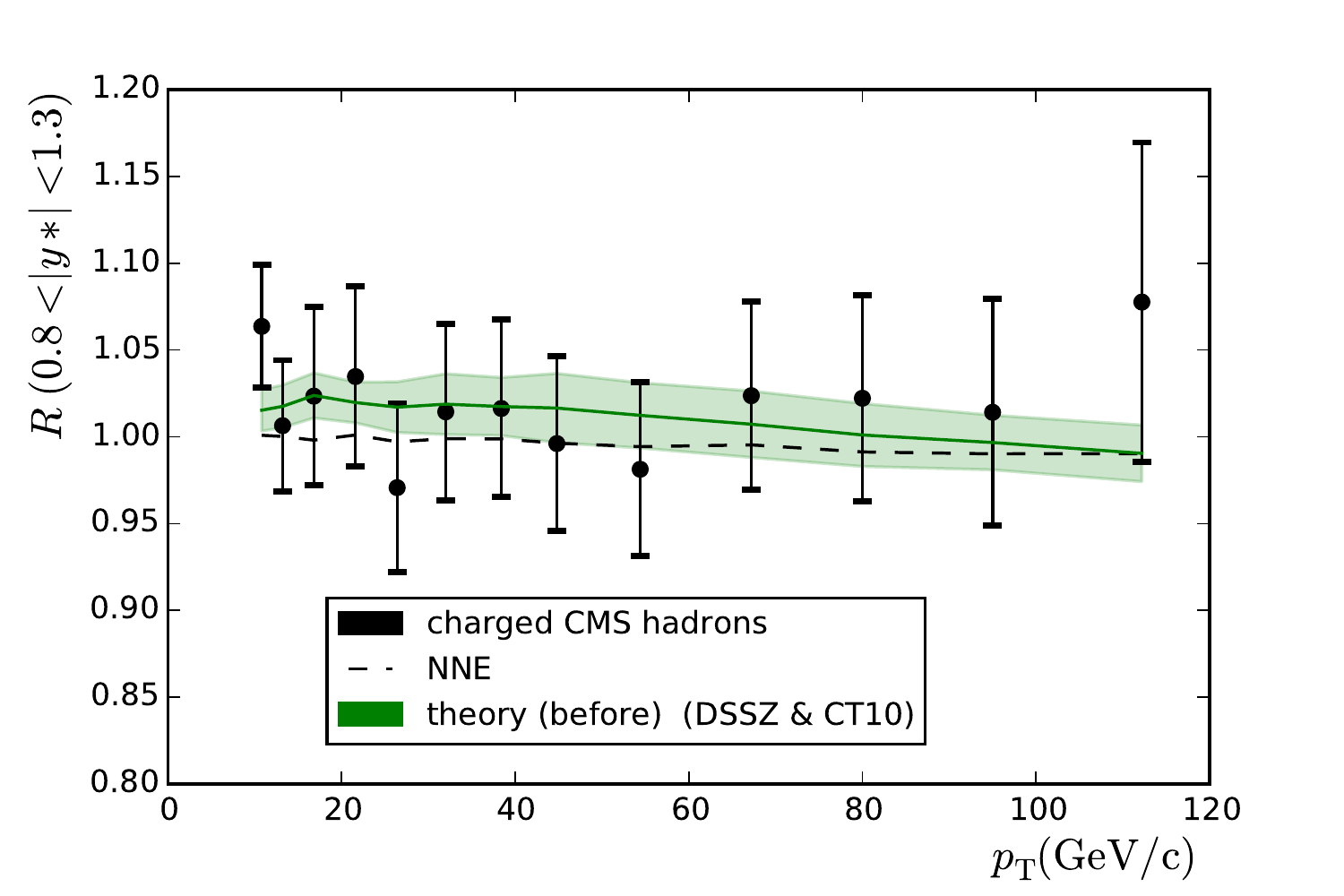}
\includegraphics[width=0.5\textwidth]{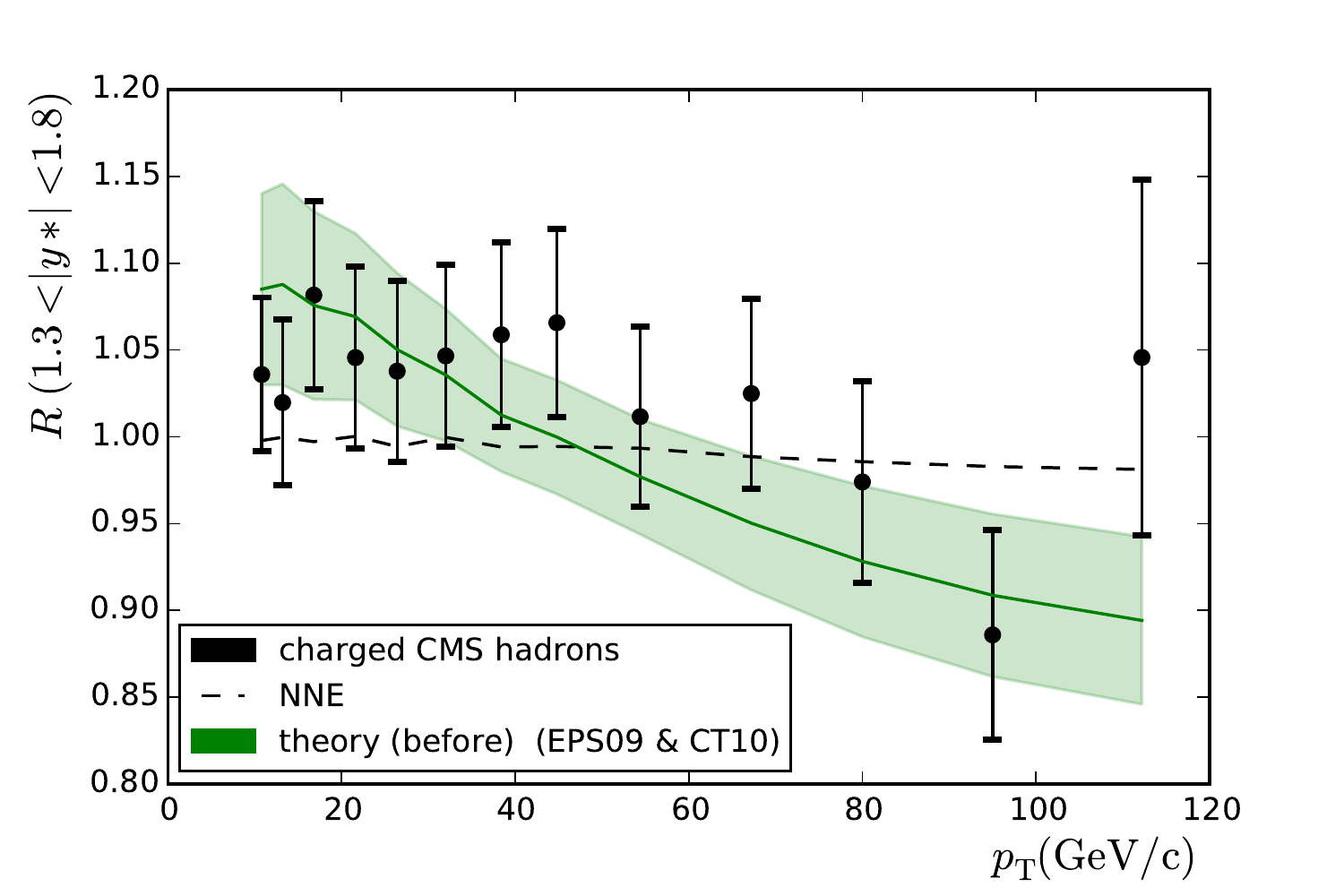} \hspace{-0.5cm}
\includegraphics[width=0.5\textwidth]{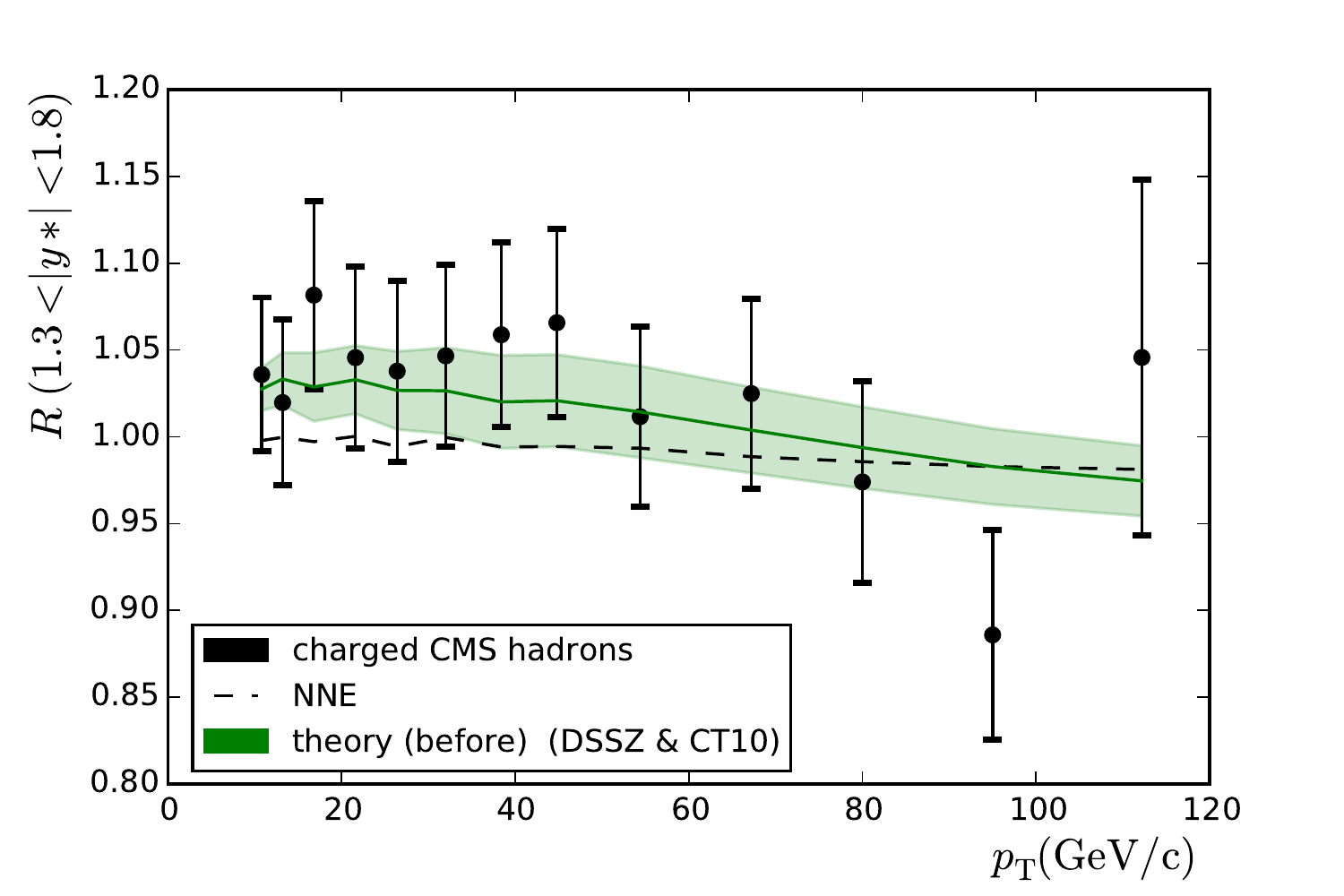}
\caption{Backward-to-forward ratios for charged-hadron production measured by the CMS collaboration. The theoretical curves were computed with EPS09 (left-hand plots) and DSSZ (right-hand plots).}
\label{fig:hadrons_cms}
\end{figure}

The use of CMS data \cite{CMS:2013cka} poses another problem since it is known that, at high-$p_{\rm T}$, the data show a 40\% enhancement that cannot currently be described by any theoretical model. However, it has been noticed that the forward-to-backward ratios are nevertheless more or less consistent with the expectations. While it is somewhat hazardous to use data in this way, we do so anyway hoping that whatever causes the high-$p_{\rm T}$ anomaly cancels in ratios. A comparison between these data and EPS09/DSSZ calculations is shown in Figure~\ref{fig:hadrons_cms} and the values of $\chi^2$ are listed in Table~\ref{tab:had} (left-hand column). These data have a tendency to favour the calculations with DSSZ but with $\chi^2/N_{\rm data}$ being absurdly low.





\begin{figure}
\centering
\includegraphics[width=0.5\textwidth]{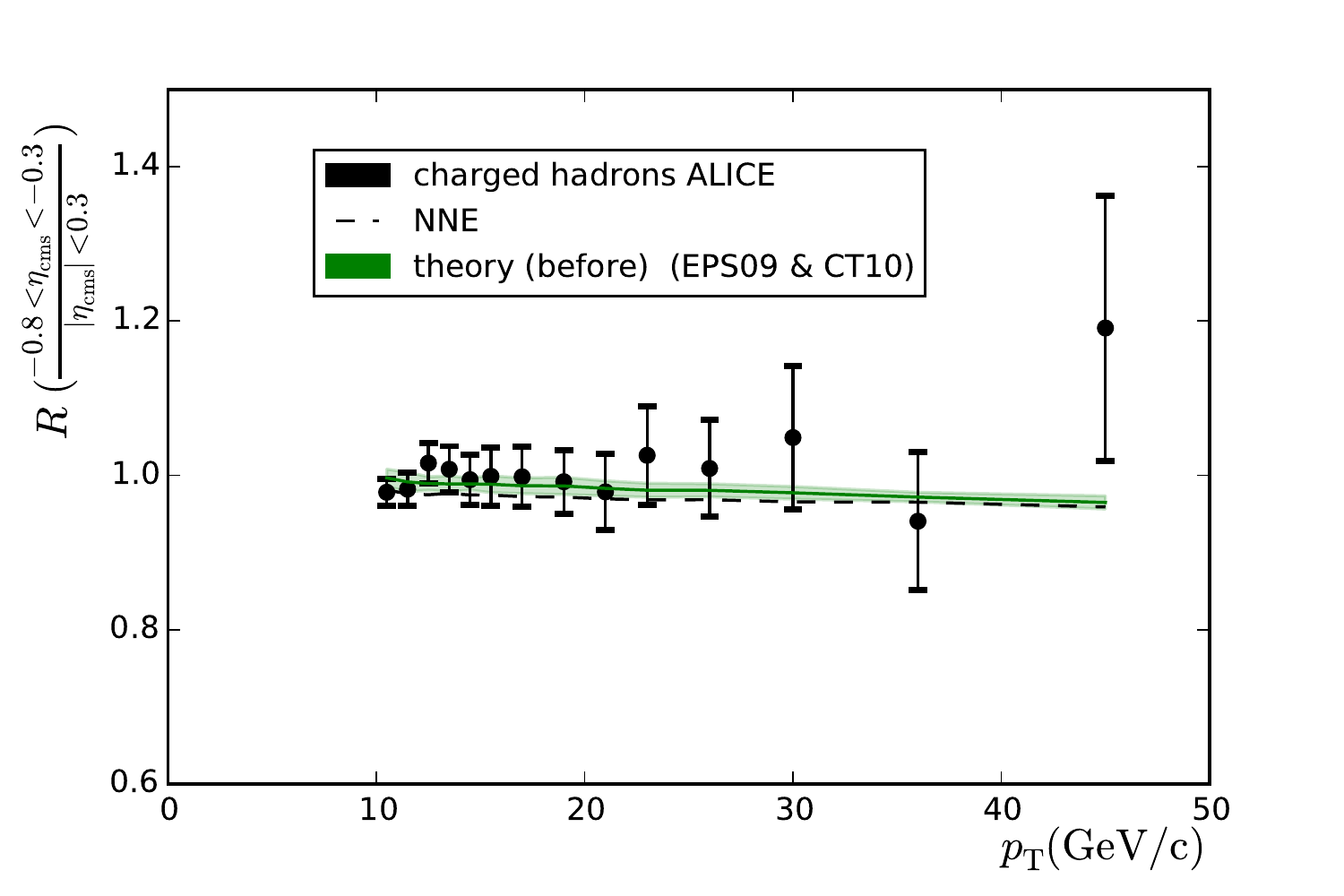} \hspace{-0.5cm}
\includegraphics[width=0.5\textwidth]{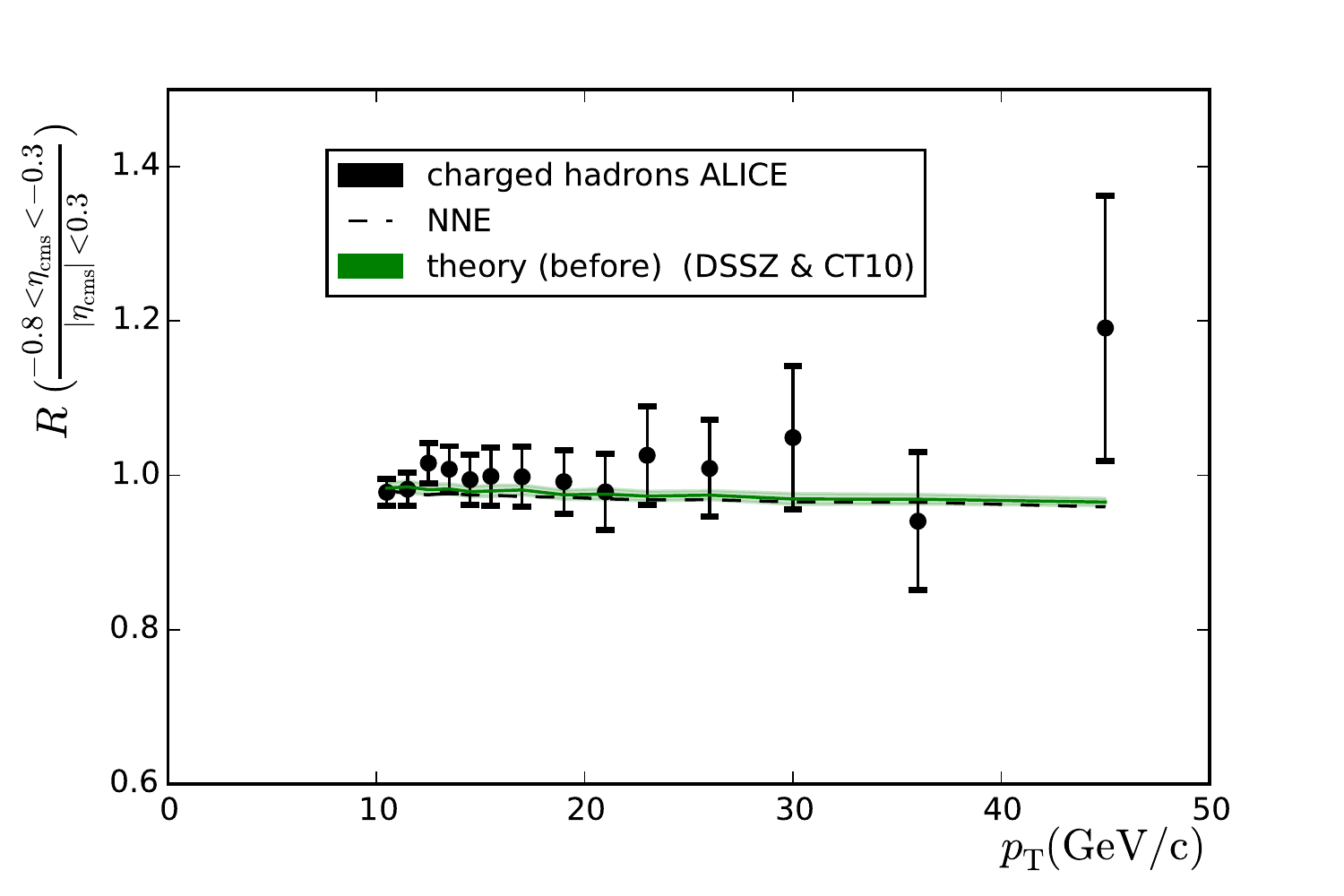}
\includegraphics[width=0.5\textwidth]{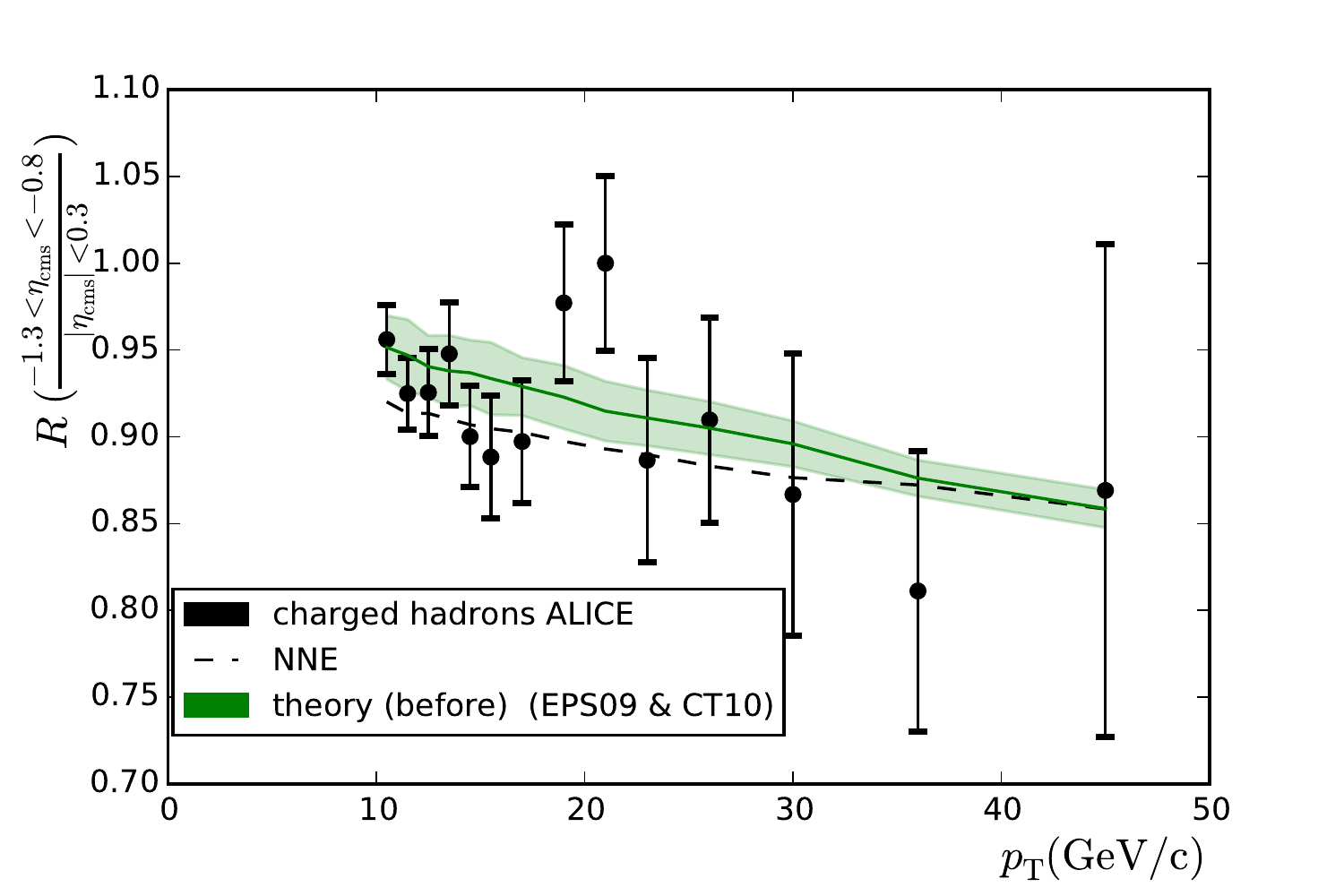} \hspace{-0.5cm}
\includegraphics[width=0.5\textwidth]{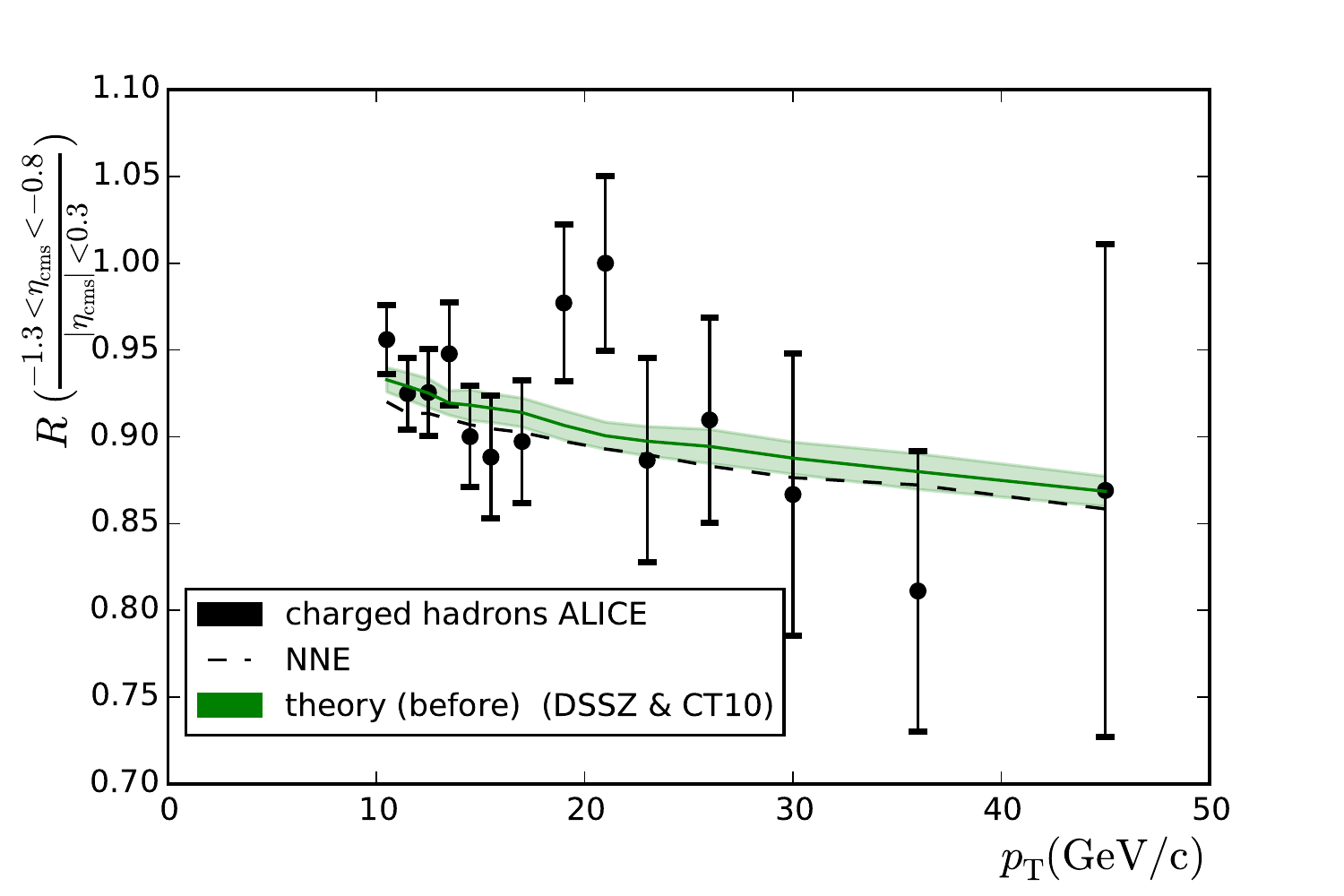}
\caption{Backward-to-central ratios of charged hadron production measured by the ALICE collaboration compared to calculations with EPS09 (left-hand panel) and DSSZ (right-hand panel).}
\label{fig:hadrons_alice}
\end{figure}

The ALICE collaboration \cite{Abelev:2014dsa} took data data relatively close to the central region and the data are available as backward-to-central ratios $A_{\rm B/C}$ 
\begin{equation}
A_{\rm B/C} =\frac{d\sigma({\rm backward})/dp_{\rm T}}{d\sigma(|\eta_{\rm c.m.}| < 0.3)/dp_{\rm T}} \, ,
\label{eq:btoc_ratio}
\end{equation}
with \emph{backward} comprising the intervals $-1.3 < \eta_{\rm c.m.} < -0.8$ and $-0.8 < \eta_{\rm c.m.} < -0.3$. A theory-to-data comparison is shown in Figure~\ref{fig:hadrons_alice} and the corresponding $\chi^2$s are in Table~\ref{tab:had} (middle column). The data appear to slightly favour the use of EPS09/DSSZ but the $\chi^2/N_{\rm data}$ remain, again, always very low.

\begin{figure}
\centering
\includegraphics[width=0.5\textwidth]{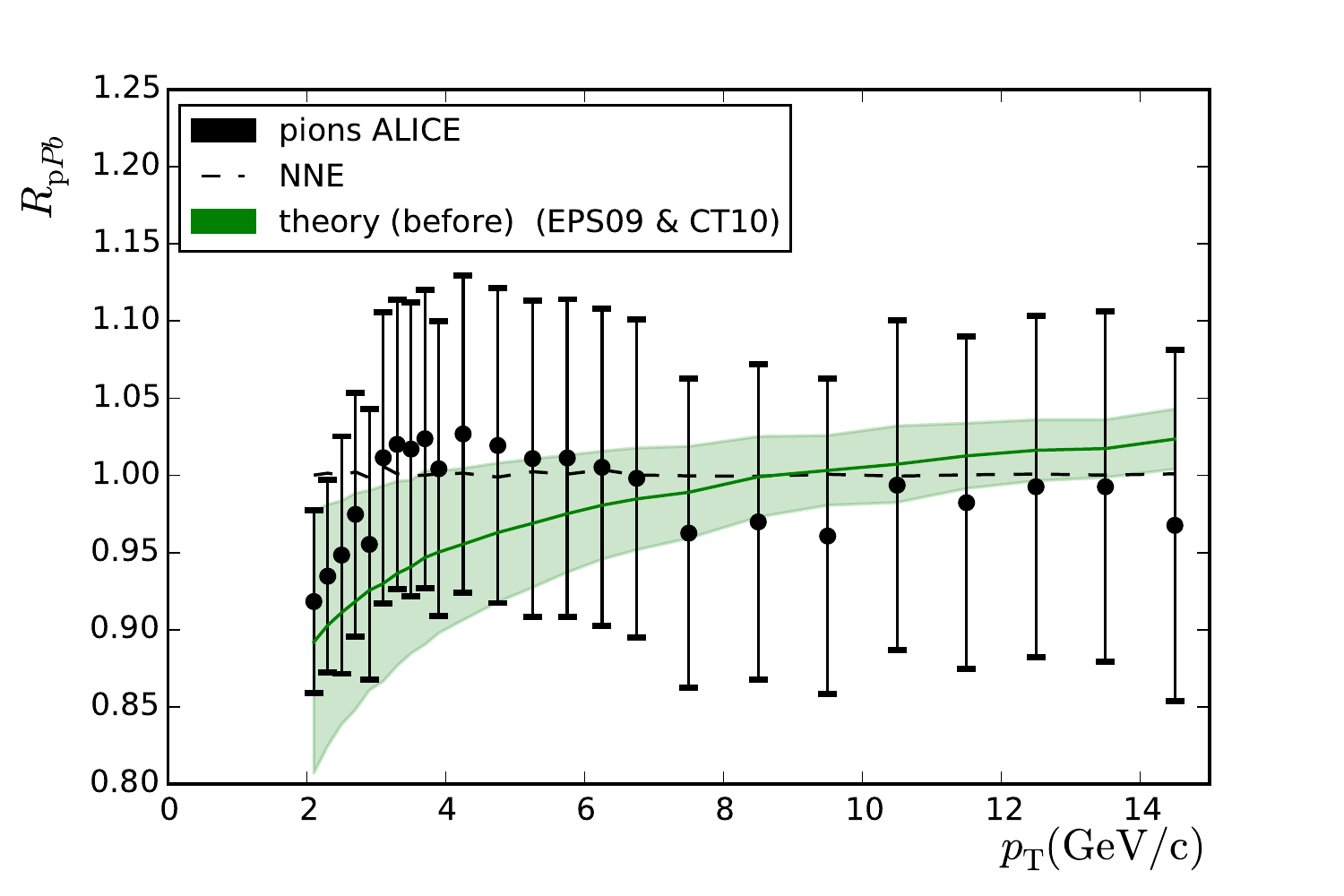} \hspace{-0.5cm}
\includegraphics[width=0.5\textwidth]{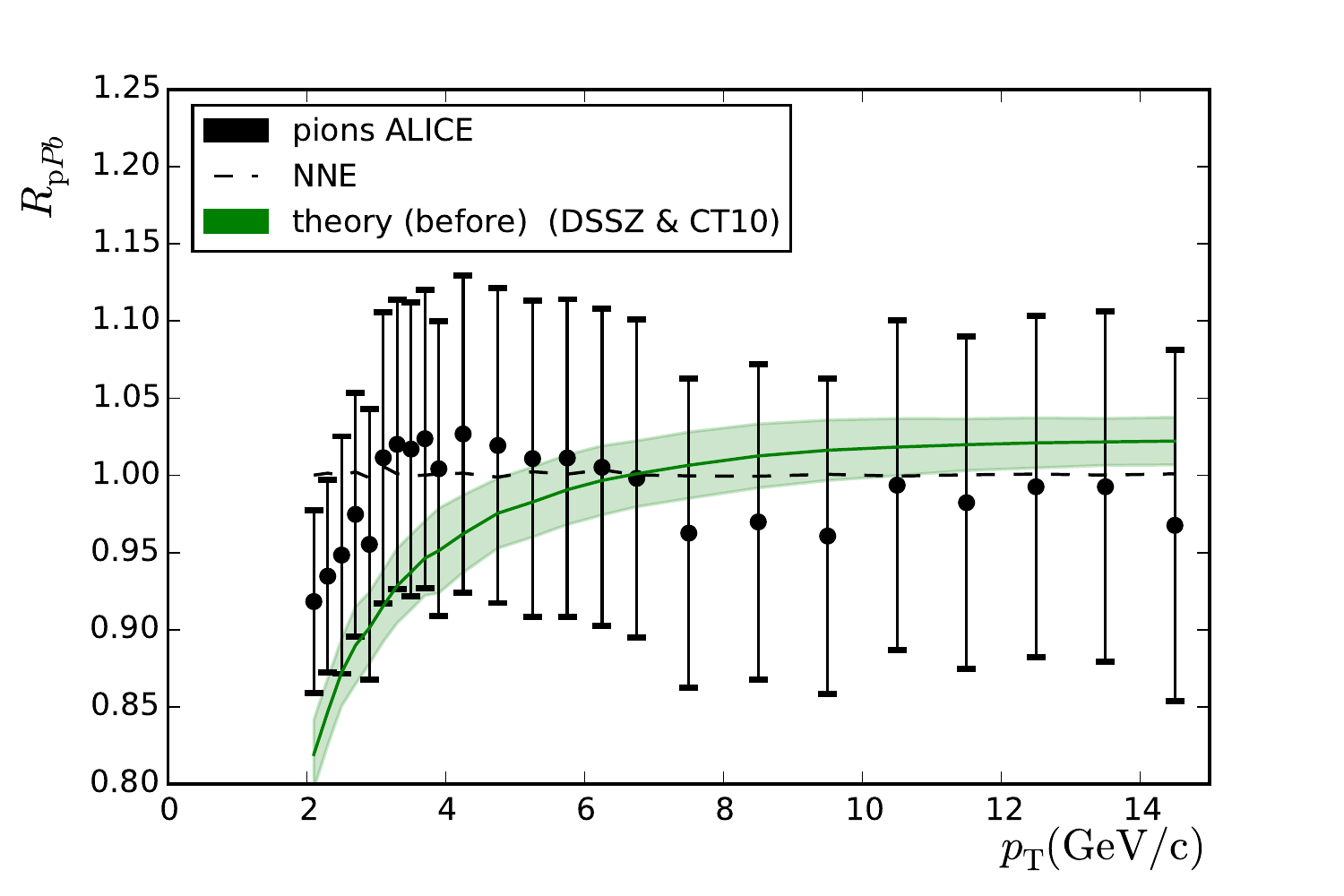}
\caption{
Ratio of minimum-bias $\pi^{+}+\pi^{-}$ production in p-Pb and the same observable in p-p collisions measured by the ALICE collaboration. Theoretical values and uncertainties were calculated with EPS09 (left-hand panel) and DSSZ (right-hand panel).}
\label{fig:pions_alice}
\end{figure}

Finally, we consider the preliminary pion data ($\pi^{+}+\pi^{-}$) shown by ALICE \cite{Richer:2015vqa}. In this case the measurement was performed only in the $|y|<0.5$ region so no $A_{\rm F/B}$ or any similar quantity could be constructed. For this reason we had to resort to the use of $R_{\rm pPb}$ ratio which involves a $6\%$ normalisation uncertainty.\footnote{Here we have deliberately ignored the normalization uncertainty --- even by doing so the obtained values of $\chi^2/N_{\rm data}$ are unrealistically small.} A comparison between data and theory before the reweighting can be seen in Figure~\ref{fig:pions_alice} and the values of $\chi^2$ are in Table~\ref{tab:had} (right-hand column). 

The very low values of $\chi^2/N_{\rm data}$ attained in these three measurements indicate that the uncertainties have been overestimated and these data are doomed to have a negligible constraining power --- notice that the uncertainties are dominated by the systematic errors which we add in quadrature with the statistical ones, in absence of a better experimental information.


\begin{table}[h!]
\begin{center}
\caption{
As Table~\ref{tab:w} but for charged particles.
}
\label{tab:had}
\begin{tabular}{cccc}
\\
${\rm PDF+nPDF}$ & $h^{+}+h^{-}_{\rm CMS}$ (39) & $h^{+}+h^{-}_{\rm ALICE}$ (28) & $\pi^{+}+\pi^{-}_{\rm ALICE}$ (24) \\
\hline
\hline
CT10+DSSZ       & 15.224 & 17.761 & 12.842 \\
CT10+EPS09      & 29.837 & 17.067 & 6.398 \\
CT10 only       & 24.075 & 23.249 & 4.644 \\
MSTW2008+DSSZ   & 15.709 & 16.970 & 16.274 \\
MSTW2008+EPS09  & 29.151 & 16.537 & 5.863 \\
MSTW2008 only   &  24.328 & 21.948 & 4.701
\end{tabular}
\end{center}
\end{table}


\section{Implications for nPDFs}\label{sec:results}

The comparisons presented in the previous section demonstrate that many of the considered data (CMS W, CMS Z, ATLAS Z, CMS dijet) show sensitivity to the nuclear PDFs while others (ALICE W, ATLAS jets, CMS hadrons, ALICE hadrons, ALICE pions) remain inconclusive. Some of the considered observables (ATLAS jets, CMS hadrons) are also known to pose issues that are not fully understood, so the comparisons presented here should be taken as indicative. The most stringent constraints are provided by the  CMS dijet measurements, which alone would rule out all but EPS09. However, upon summing all the $\chi^2$'s from the different measurements, this easily gets buried under the other data. This is evident from the total values of $\chi^2/N_{\rm data}$ shown in Table~\ref{tab:numbers} (upper part), as considering all the data it would look like all the PDF combinations were in agreement with the data ($\chi^2/N_{\rm data} \sim 1 $). However, excluding one of the dubious data sets (ATLAS jets) for which the number of data is large but $\chi^2/N_{\rm data}$ very small, the differences between different PDFs grow, see the lower part of Table~\ref{tab:numbers}. The effective number of replicas remains always quite high. The reason for the high $N_{\rm eff}$ is that the variation of the total $\chi^2$ within a given set of nPDFs (that is, the variation among the error sets) is small even if some of the data sets are not properly described at all (in particular, CMS dijets with DSSZ). Thus, $N_{\rm eff}$ alone should not be blindly used to judge whether a reanalysis is required.

\begin{table}[h!]
\begin{center}
\caption{Values of the $\chi^{2}/N_{\rm data}$ before and after the reweighting, and the effective number of remaining replicas $N_{\rm eff}$. The upper part corresponds to considering all the data (165 data points in total) and the lower part excluding the ATLAS jet measurements (130 data points in total).}
\label{tab:numbers}
\begin{tabular}{cccccc}
All data & ${\rm PDF+nPDF}$ & $\chi^{2}_{\rm original}$ & $\chi^{2}_{\rm reweighted}$ & $N_{\rm eff}$ \\
\hline
\hline
& CT10+DSSZ       & 1.074 & 1.016 & 9044 \\
& CT10+EPS09      & 0.674 & 0.632 & 8657 \\
& MSTW2008+DSSZ   & 0.876 & 0.826 & 9128 \\
& MSTW2008+EPS09  & 0.649 & 0.583 & 8585 \\
& CT10 only       & 1.425  &   -   &  -   \\
& MSTW2008 only   & 1.138  &   -   &  -   \\
\\
Excluding ATLAS jets
& ${\rm PDF+nPDF}$ & $\chi^{2}_{\rm original}$ & $\chi^{2}_{\rm reweighted}$ & $N_{\rm eff}$ \\
\hline
\hline
& CT10+DSSZ       & 1.277 &   1.199   & 9014 \\
& CT10+EPS09      & 0.679 &   0.638   & 8706 \\
& MSTW2008+DSSZ   & 1.023 &   0.957   & 9107 \\
& MSTW2008+EPS09  & 0.652 &   0.589   & 8697 \\
& CT10 only       & 1.734 &   -   &  -   \\
& MSTW2008 only   & 1.369 &   -   &  -
\end{tabular}
\end{center}
\end{table}

\begin{figure}
\centering
\includegraphics[width=0.5\textwidth]{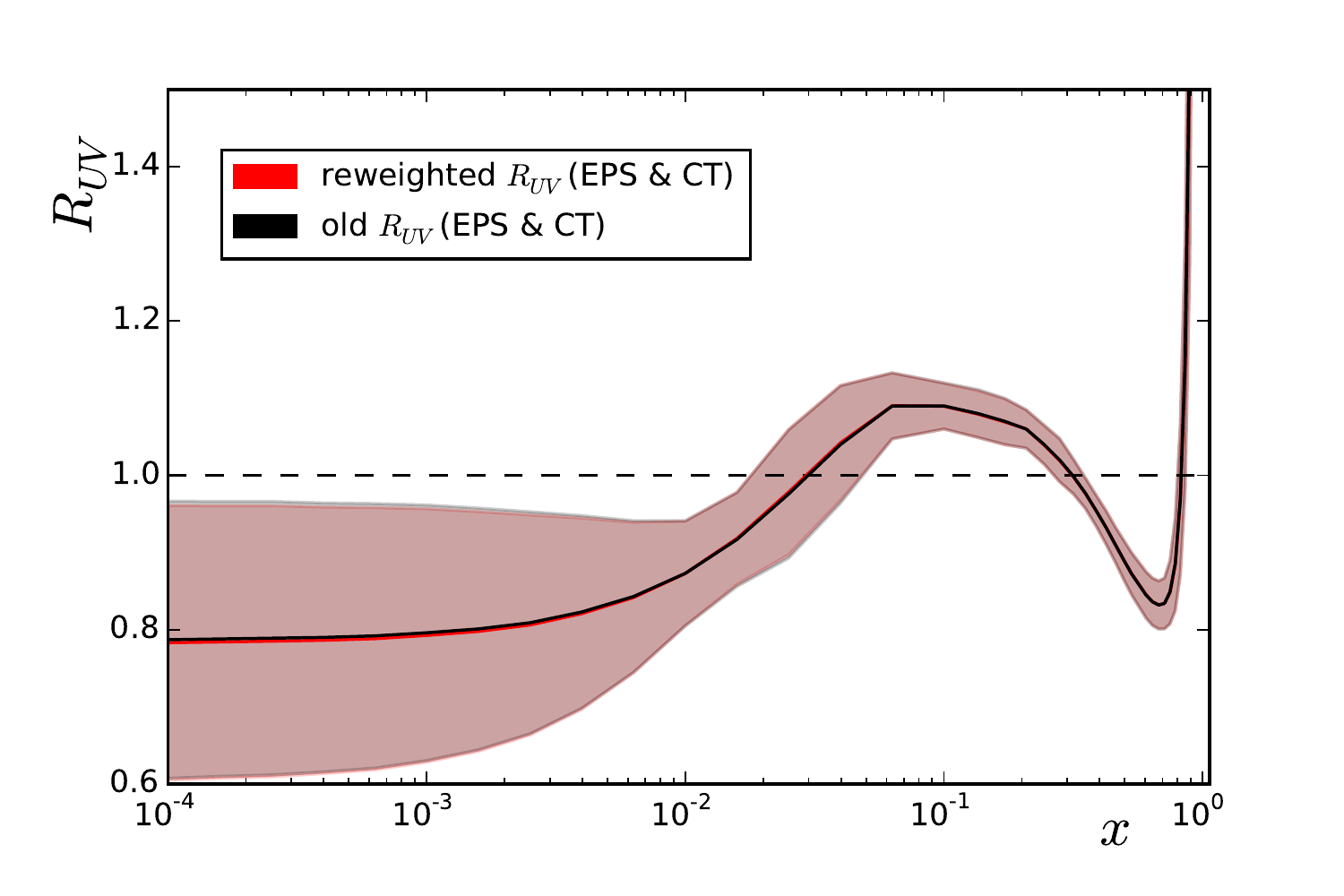} \hspace{-0.5cm}
\includegraphics[width=0.5\textwidth]{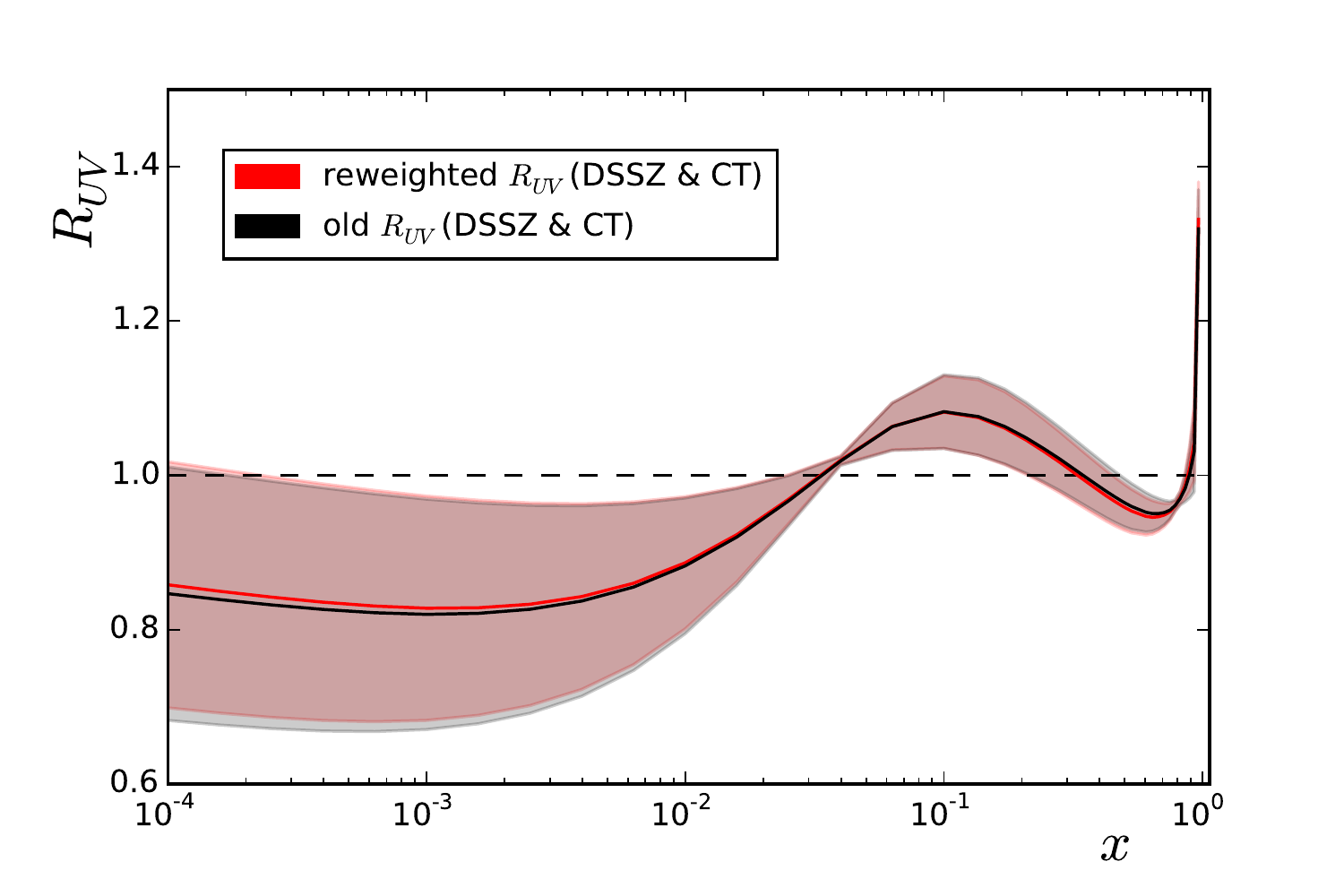}
\includegraphics[width=0.5\textwidth]{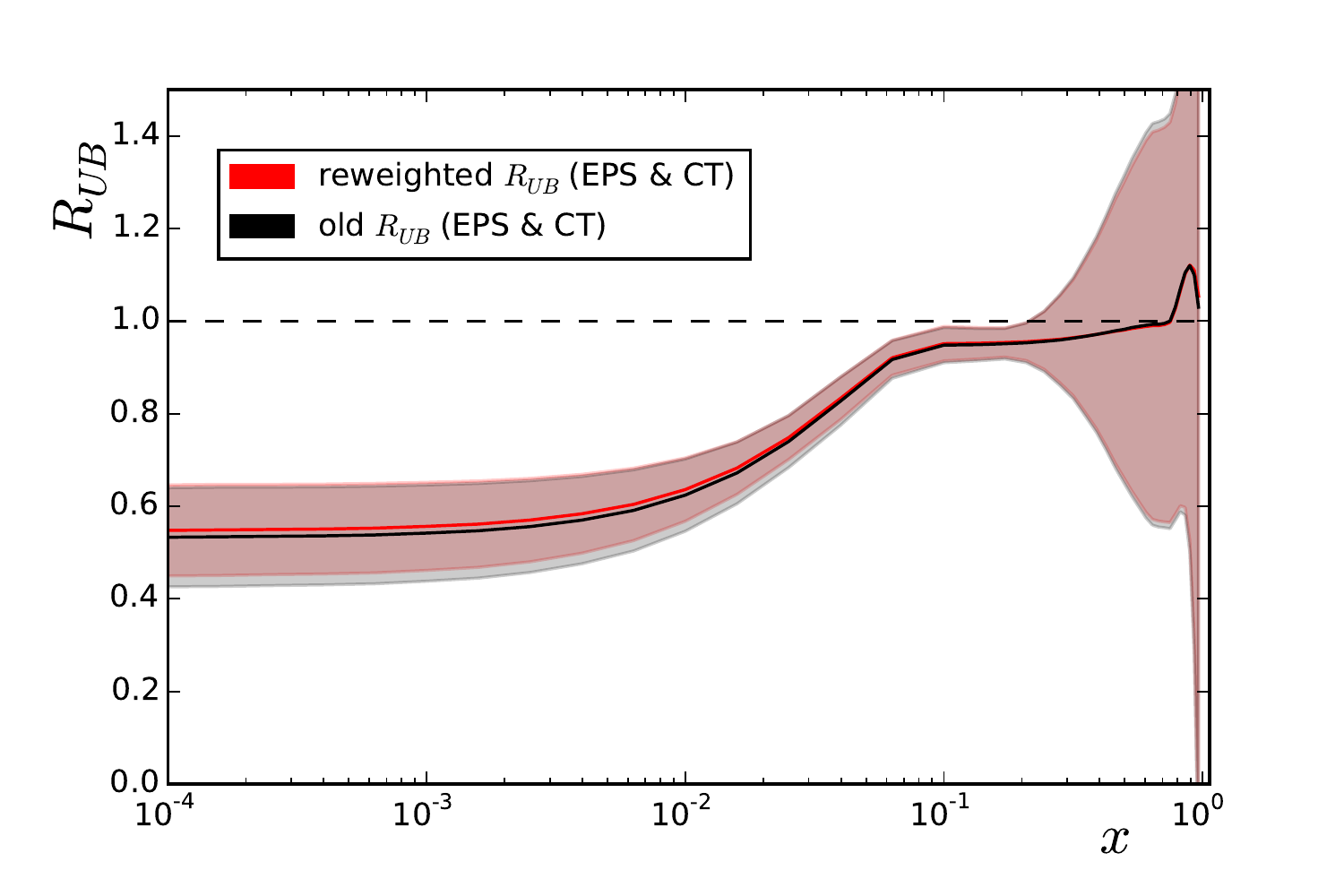} \hspace{-0.5cm}
\includegraphics[width=0.5\textwidth]{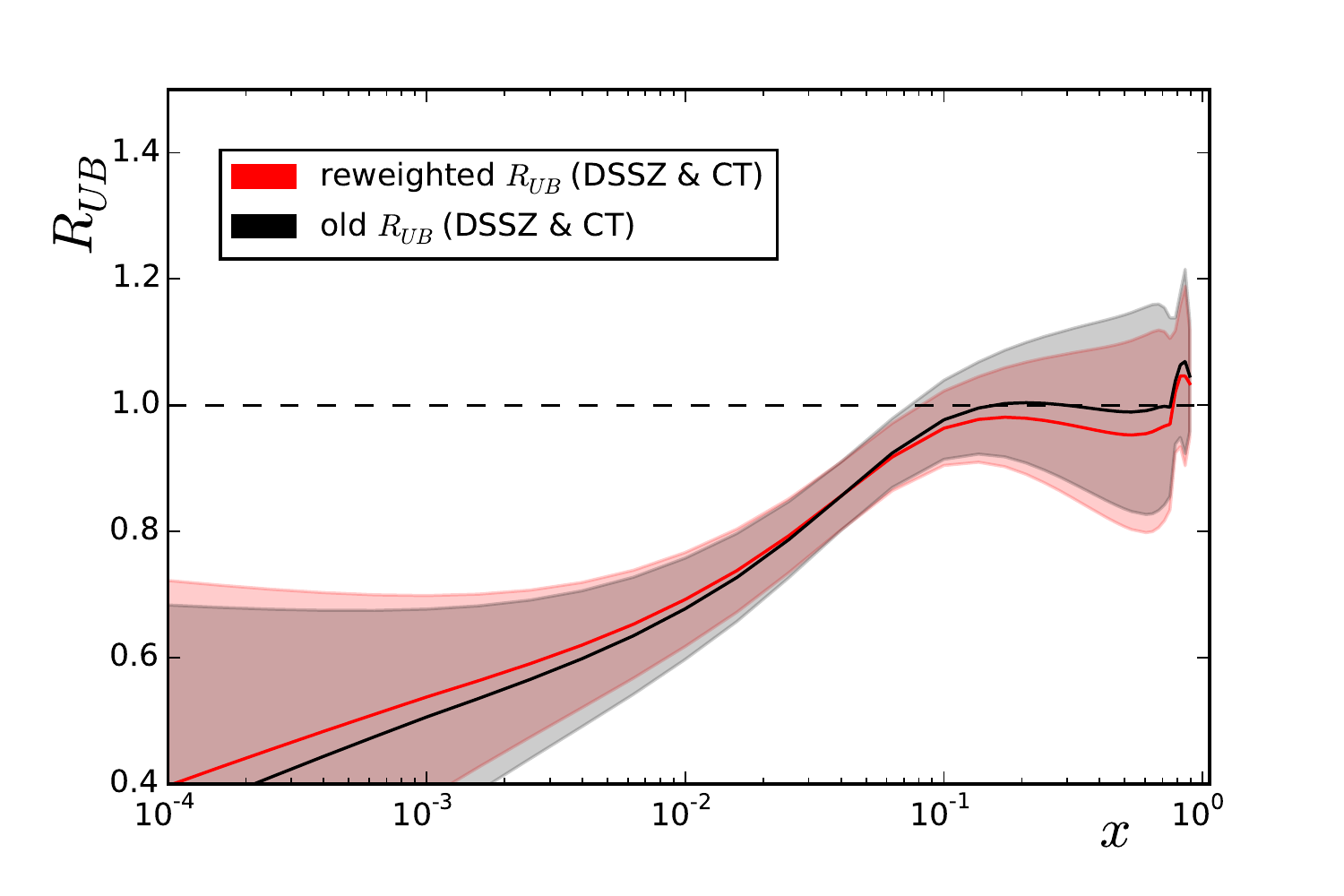}
\includegraphics[width=0.5\textwidth]{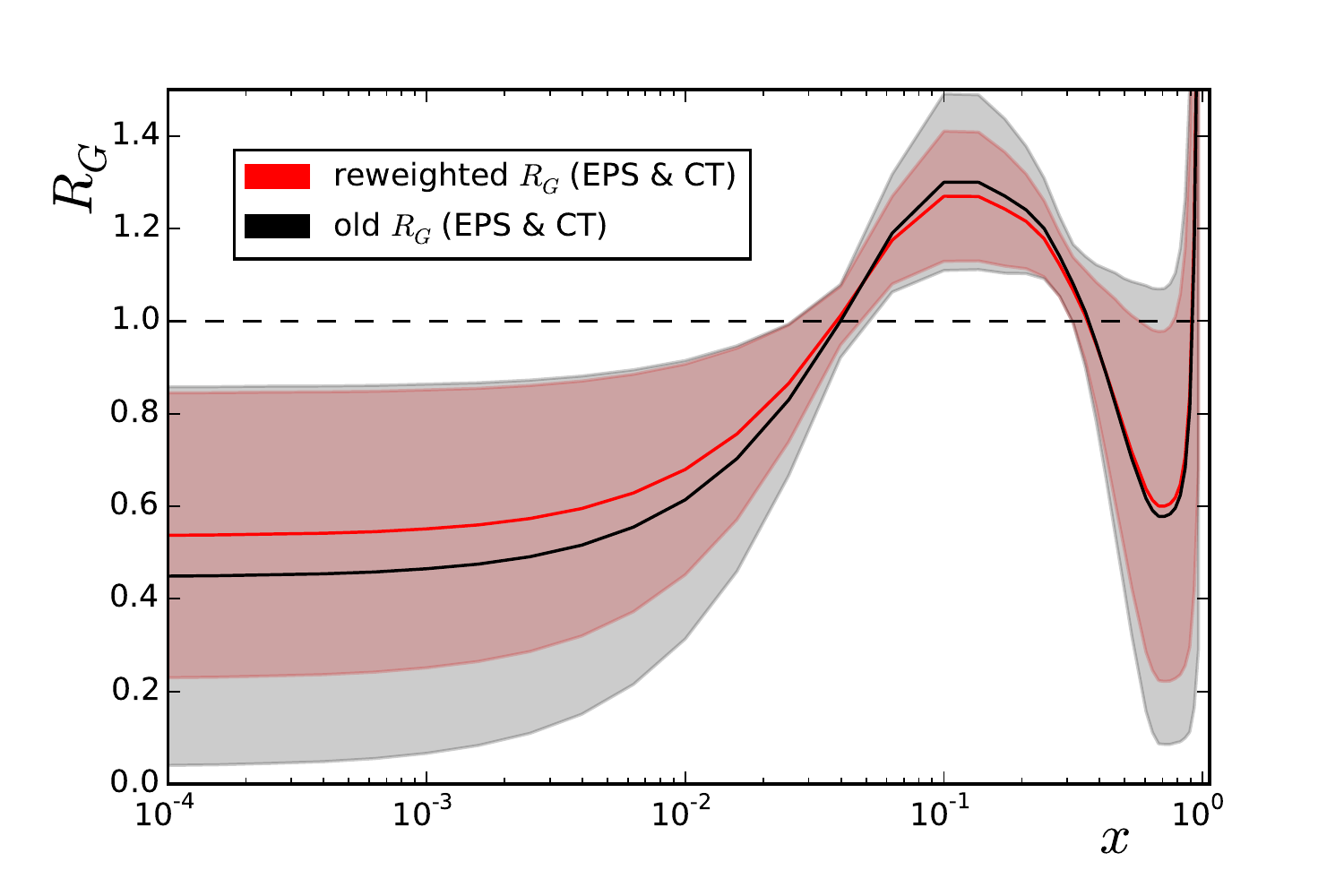} \hspace{-0.5cm}
\includegraphics[width=0.5\textwidth]{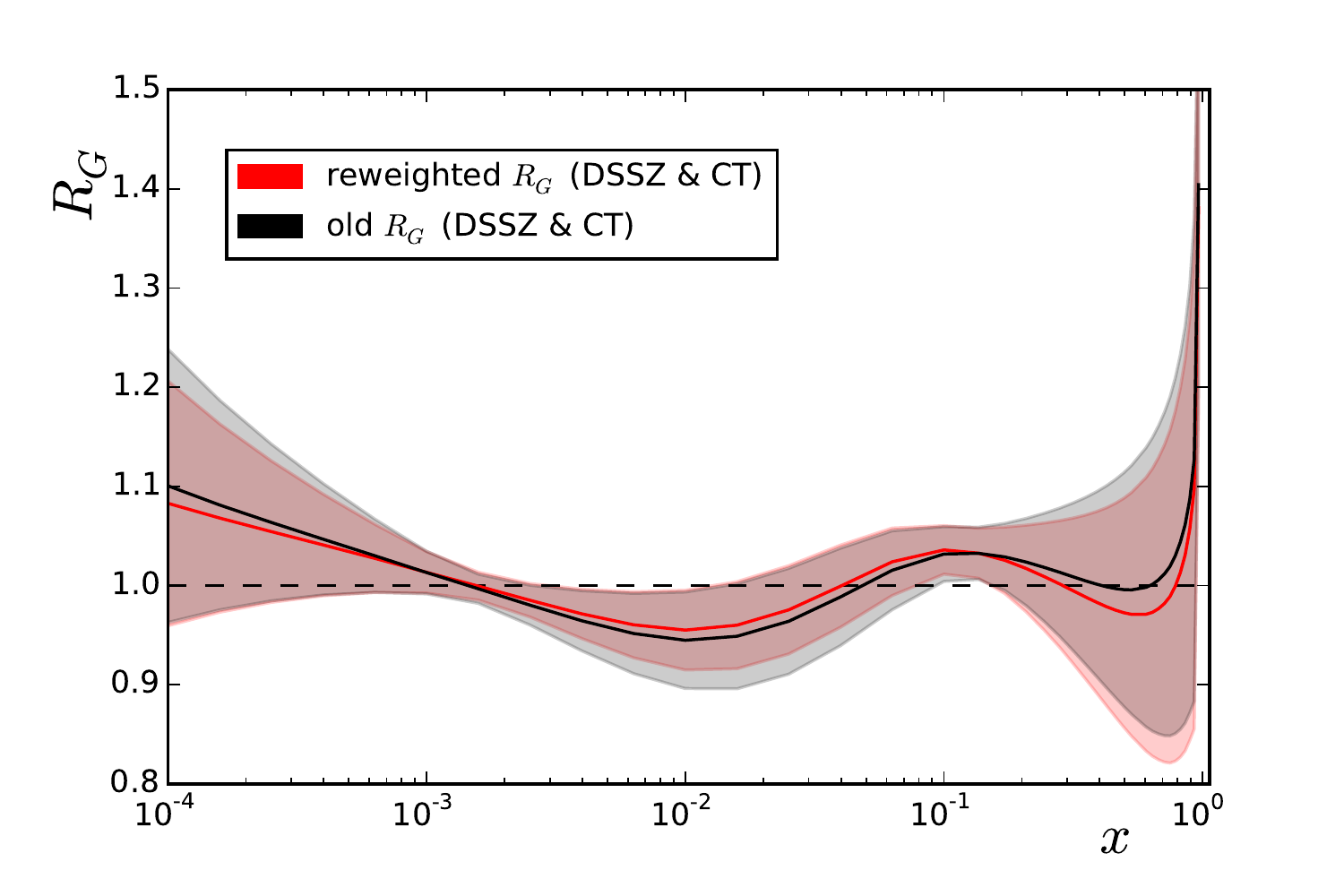}
\caption{Impact of the LHC Run I data on the nPDFs of EPS09 (left) and DSSZ (right) before (black/grey) and after the reweighting (red/light red), for valence (upper panels), sea (middle panels) and gluon (lower panels) distributions at $Q^{2}=1.69 \, {\rm GeV}^{2}$, except the DSSZ gluons that are plotted at $Q^{2}=2 \, {\rm GeV}^{2}$.
}
\label{fig:pdfs_reweighted}
\end{figure} 

Given the tiny improvements in reweighted $\chi^2$ values one expects no strong modifications to be induced in the nPDFs either. Indeed, the only noticeable effect, as can be seen in Figure~\ref{fig:pdfs_reweighted}, is in the EPS09 gluons for which the CMS dijet data place new constraints \cite{Paukkunen:2014pha}.\footnote{These are results using all the data, including those whose consistency is in doubt.} It should be recalled that, for technical reasons, in the EPS09 analysis the RHIC pion data were given a rather large additional weight and they still overweight the $\chi^2$ contribution coming from the dijets. In a fit with no extra weights the dijet data would, on the contrary, give a larger contribution than the RHIC data. Therefore these data will have a different effect that what Figure~\ref{fig:pdfs_reweighted} would indicate. In the case of DSSZ the assumed functional form is not flexible enough to accommodate the dijet data and in practice nothing happens upon performing the reweighting. However, it is evident that these data will have a large impact on the DSSZ gluons if an agreement is required (see Figure~\ref{fig:di-jets_cms}), so a refit appears mandatory. 

The impact of the LHC p-Pb data is potentially higher than what is found here also since, within our study, it is impossible to say anything concerning the constraints that these data may provide for the flavour separation of the nuclear PDFs, which again calls for a refit. Another issue is the form of the fit functions whose rigidity especially at small $x$ significantly underestimates the true uncertainty. In this sense, our study should be seen merely as a preparatory work towards nPDFs analyses including LHC data. More data p-Pb will also still appear (at least CMS inclusive jets, W production from ATLAS) and many of the data sets used here are only preliminary. 

\section{Summary}\label{sec:summary}
In the present work we have examined the importance of PDF nuclear modifications in describing some p-Pb results from Run I at the LHC, and the impact that the considered data  have  on the EPS09 and DSSZ global fits of nPDFs. We have found that while some data clearly favors the considered sets of nuclear PDFs, some sets are also statistically consistent  with just proton PDFs. In this last case abnormally small values of $\chi^2/N_{\rm data}$ are obtained, however. The global picture therefore depends on what data sets are being considered. We have chosen to use, in our analysis, most of the available data from the p-Pb run, it should, however, be stressed that some of the considered data sets are suspicious in the sense that  unrealistically small values of $\chi^2/N_{\rm data}$ are obtained and these sets, as we have shown, can easily twist the overall picture. Incidentally, these sets are the ones that have smallest $\chi^2$ when no nuclear effects in PDFs are included. The small values of $\chi^2/N_{\rm data}$ are partly related to unknown correlations between the systematic uncertainties of the data but also, particularly in the case of ALICE pions, presumably to the additional uncertainty added to the interpolated p-p baseline. The p-p reference data at $\sqrt{s}=5.02 \, {\rm TeV}$, recently recorded at the LHC, may eventually improve this situation.

The considered data are found to have only a  mild impact on the EPS09 and DSSZ nPDFs. This does not, however, necessarily mean that these data would be  useless. Indeed, they may facilitate to relax some rather restrictive assumptions made in the fits. An obvious example is the functional form for DSSZ gluon modification which does not allow for a similar gluon antishadowing as the EPS09 fit functions. This leads to a poor description of the CMS dijet data by DSSZ that the reweighting (being restricted to all assumptions made in the original analysis) cannot cure. Thus, in reality, these data are likely to have a large impact. In general, these new LHC data may allow to implement more flexibility into the fit functions and also to release restrictions related to the flavour dependence of the quark nuclear effects. Also, the EPS09 analysis used an additional weight to emphasise the importance of the data set (neutral pions at RHIC) sensitive to gluon nPDF. Now, with the use of the new LHC data, such artificial means are likely to be unnecessary. Therefore, for understanding the true significance of these data, new global fits including these and upcoming data are thus required.

Hence, both theoretical and experimental efforts, as explained above, are required to fully exploit the potentiality of both already done and future p-Pb runs at the LHC for constraining the nuclear modifications of parton densities.

\section*{Acknowledgments}
We thank E. Chapon and A. Zsigmond,  Z. Citron, and M. Ploskon, for their help with the understanding of the CMS, ATLAS and ALICE data respectively. This research was supported by the European Research Council grant HotLHC ERC-2011-StG-279579; the People Programme (Marie Curie Actions) of the European Union's Seventh Framework Programme FP7/2007-2013/ under REA grant agreement \#318921 (NA); Ministerio de Ciencia e Innovaci\'on of Spain under project FPA2014-58293-C2-1-P; Xunta de Galicia (Conseller\'{\i}a de Educaci\'on) --- the group is part of the Strategic Unit AGRUP2015/11.

\appendix

\end{document}